\documentclass[aps,prc,twocolumn,amsfonts,superscriptaddress,showpacs,showkeys,nofootinbib,floatfix]{revtex4-1}

\usepackage{url}
\usepackage{color}
\usepackage{graphics,epsfig}
\usepackage{verbatim}

\usepackage{amsmath}
\usepackage{amssymb}
\usepackage{amstext}
\usepackage{mathrsfs}
\usepackage{latexsym}
\usepackage{subfigure}
\usepackage{wasysym}
\usepackage{fancyhdr}

\usepackage{bm}

\newcommand {\be} {\begin{equation}}
\newcommand {\ba} {\begin{eqnarray}}
\newcommand {\ee} {\end{equation}}
\newcommand {\ea} {\end{eqnarray}}

\begin{document}
\hyphenation{RCSLACPOL}

\newcommand*{\ODU}{Old Dominion University, Norfolk, Virginia 23529}
\newcommand*{\ODUindex}{23}
\affiliation{\ODU}
\newcommand*{\ANL}{Argonne National Laboratory, Argonne, Illinois 60439}
\newcommand*{\ANLindex}{1}
\affiliation{\ANL}
\newcommand*{\ASU}{Arizona State University, Tempe, Arizona 85287-1504}
\newcommand*{\ASUindex}{1}
\affiliation{\ASU}
\newcommand*{\CSUDH}{California State University, Dominguez Hills, Carson, CA 90747}
\newcommand*{\CSUDHindex}{2}
\affiliation{\CSUDH}
\newcommand*{\CANISIUS}{Canisius College, Buffalo, NY}
\newcommand*{\CANISIUSindex}{3}
\affiliation{\CANISIUS}
\newcommand*{\CMU}{Carnegie Mellon University, Pittsburgh, Pennsylvania 15213}
\newcommand*{\CMUindex}{4}
\affiliation{\CMU}
\newcommand*{\CUA}{Catholic University of America, Washington, D.C. 20064}
\newcommand*{\CUAindex}{6}
\affiliation{\CUA}
\newcommand*{\SACLAY}{CEA, Centre de Saclay, Irfu/Service de Physique Nucl\'eaire, 91191 Gif-sur-Yvette, France}
\newcommand*{\SACLAYindex}{5}
\affiliation{\SACLAY}
\newcommand*{\UCONN}{University of Connecticut, Storrs, Connecticut 06269}
\newcommand*{\UCONNindex}{6}
\affiliation{\UCONN}
\newcommand*{\FU}{Fairfield University, Fairfield CT 06824}
\newcommand*{\FUindex}{11}
\affiliation{\FU}
\newcommand*{\FIU}{Florida International University, Miami, Florida 33199}
\newcommand*{\FIUindex}{7}
\affiliation{\FIU}
\newcommand*{\FSU}{Florida State University, Tallahassee, Florida 32306}
\newcommand*{\FSUindex}{8}
\affiliation{\FSU}
\newcommand*{\GWUI}{The George Washington University, Washington, DC 20052}
\newcommand*{\GWUIindex}{9}
\affiliation{\GWUI}
\newcommand*{\ISU}{Idaho State University, Pocatello, Idaho 83209}
\newcommand*{\ISUindex}{10}
\affiliation{\ISU}
\newcommand*{\INFNFE}{INFN, Sezione di Ferrara, 44100 Ferrara, Italy}
\newcommand*{\INFNFEindex}{11}
\affiliation{\INFNFE}
\newcommand*{\INFNFR}{INFN, Laboratori Nazionali di Frascati, 00044 Frascati, Italy}
\newcommand*{\INFNFRindex}{12}
\affiliation{\INFNFR}
\newcommand*{\INFNGE}{INFN, Sezione di Genova, 16146 Genova, Italy}
\newcommand*{\INFNGEindex}{13}
\affiliation{\INFNGE}
\newcommand*{\INFNRO}{INFN, Sezione di Roma Tor Vergata, 00133 Rome, Italy}
\newcommand*{\INFNROindex}{14}
\affiliation{\INFNRO}
\newcommand*{\INFNTUR}{INFN, Sezione di Torino, 10125 Torino, Italy}
\newcommand*{\INFNTURindex}{15}
\affiliation{\INFNTUR}
\newcommand*{\ORSAY}{Institut de Physique Nucl\'eaire, CNRS/IN2P3 and Universit\'e Paris Sud, Orsay, France}
\newcommand*{\ORSAYindex}{16}
\affiliation{\ORSAY}
\newcommand*{\ITEP}{Institute of Theoretical and Experimental Physics, Moscow, 117259, Russia}
\newcommand*{\ITEPindex}{17}
\affiliation{\ITEP}
\newcommand*{\JMU}{James Madison University, Harrisonburg, Virginia 22807}
\newcommand*{\JMUindex}{18}
\affiliation{\JMU}
\newcommand*{\KNU}{Kyungpook National University, Daegu 702-701, Republic of Korea}
\newcommand*{\KNUindex}{19}
\affiliation{\KNU}
\newcommand*{\UNH}{University of New Hampshire, Durham, New Hampshire 03824-3568}
\newcommand*{\UNHindex}{20}
\affiliation{\UNH}
\newcommand*{\NSU}{Norfolk State University, Norfolk, Virginia 23504}
\newcommand*{\NSUindex}{21}
\affiliation{\NSU}
\newcommand*{\OHIOU}{Ohio University, Athens, Ohio  45701}
\newcommand*{\OHIOUindex}{22}
\affiliation{\OHIOU}
\newcommand*{\RPI}{Rensselaer Polytechnic Institute, Troy, New York 12180-3590}
\newcommand*{\RPIindex}{24}
\affiliation{\RPI}
\newcommand*{\URICH}{University of Richmond, Richmond, Virginia 23173}
\newcommand*{\URICHindex}{25}
\affiliation{\URICH}
\newcommand*{\ROMAII}{Universita' di Roma Tor Vergata, 00133 Rome Italy}
\newcommand*{\ROMAIIindex}{26}
\affiliation{\ROMAII}
\newcommand*{\MSU}{Skobeltsyn Institute of Nuclear Physics, Lomonosov Moscow State University, 119234 Moscow, Russia}
\newcommand*{\MSUindex}{27}
\affiliation{\MSU}
\newcommand*{\SCAROLINA}{University of South Carolina, Columbia, South Carolina 29208}
\newcommand*{\SCAROLINAindex}{28}
\affiliation{\SCAROLINA}
\newcommand*{\TEMPLE}{Temple University,  Philadelphia, PA 19122 }
\newcommand*{\TEMPLEindex}{29}
\affiliation{\TEMPLE}
\newcommand*{\JLAB}{Thomas Jefferson National Accelerator Facility, Newport News, Virginia 23606}
\newcommand*{\JLABindex}{30}
\affiliation{\JLAB}
\newcommand*{\UTFSM}{Universidad T\'{e}cnica Federico Santa Mar\'{i}a, Casilla 110-V Valpara\'{i}so, Chile}
\newcommand*{\UTFSMindex}{31}
\affiliation{\UTFSM}
\newcommand*{\EDINBURGH}{Edinburgh University, Edinburgh EH9 3JZ, United Kingdom}
\newcommand*{\EDINBURGHindex}{32}
\affiliation{\EDINBURGH}
\newcommand*{\GLASGOW}{University of Glasgow, Glasgow G12 8QQ, United Kingdom}
\newcommand*{\GLASGOWindex}{33}
\affiliation{\GLASGOW}
\newcommand*{\VIRGINIA}{University of Virginia, Charlottesville, Virginia 22901}
\newcommand*{\VIRGINIAindex}{34}
\affiliation{\VIRGINIA}
\newcommand*{\WM}{College of William and Mary, Williamsburg, Virginia 23187-8795}
\newcommand*{\WMindex}{35}
\affiliation{\WM}
\newcommand*{\YEREVAN}{Yerevan Physics Institute, 375036 Yerevan, Armenia}
\newcommand*{\YEREVANindex}{36}
\affiliation{\YEREVAN}
 
\newcommand*{\NOWKENTU}{Christopher Newport University, Newport News, Virginia 23606}
\newcommand*{\NOWSSI}{Spectral Sciences Inc., Burlington, MA 01803}

\newcommand*{\NOWJLAB}{Thomas Jefferson National Accelerator Facility, Newport News, Virginia 23606}
\newcommand*{\NOWODU}{Old Dominion University, Norfolk, Virginia 23529}

\title{Precise Determination of the Deuteron Spin Structure at Low to Moderate $Q^2$ with CLAS and Extraction of the Neutron Contribution}

\author{N. Guler}
\altaffiliation[Current address: ]{\NOWSSI}
\affiliation{\ODU}
\author{R.G.\ Fersch}
\altaffiliation[Current address: ]{\NOWKENTU}
\affiliation{\WM}
\author {S.E.~Kuhn} 
     \thanks{Corresponding author. Email: skuhn@odu.edu}
\affiliation{\ODU}
\author{P.~Bosted}
\affiliation{\WM}
\affiliation{\JLAB}
\author{K.A.~Griffioen}
\affiliation{\WM}
\author{C.~Keith}
\affiliation{\JLAB}
\author{R.~Minehart}
\affiliation{\VIRGINIA}
\author {Y.~Prok} 
\affiliation{\ODU}
\affiliation{\JLAB}
\author {K.P.~Adhikari} 
\affiliation{\ODU}
\author {D.~Adikaram} 
\altaffiliation[Current address: ]{\NOWJLAB}
\affiliation{\ODU}
% ...

\collaboration{The CLAS Collaboration}
     \noaffiliation
\date{\today}

\begin{abstract}

We present the final results for the deuteron spin structure
functions obtained from the full data set collected with Jefferson Lab's CLAS
in 2000-2001.
Polarized electrons with
energies of 1.6, 2.5, 4.2 and 5.8 GeV  were scattered from deuteron ($^{15}$ND$_3$)
targets, dynamically polarized along the beam direction, and detected with CLAS. From the
measured double spin asymmetry, the virtual photon absorption asymmetry $A_1^d$ and
the polarized structure function   $g_1^d$
were extracted over a wide kinematic range (0.05~GeV$^2 < Q^2 <$~5~GeV$^2$ and 
0.9~GeV~$< W <$~3~GeV). We use an unfolding procedure 
and a parametrization of the corresponding proton results
to extract from these data  the polarized structure functions
$A_1^n$ and  $g_1^n$ of the (bound) neutron, which are so far unknown in the resonance region,
$W < 2$ GeV.
We compare our final results, including several moments of the deuteron and neutron
spin structure functions,
 with various theoretical
models and expectations as well as parametrizations of the world data. The unprecedented precision
and dense kinematic coverage of these data  can aid in future extractions of
polarized parton distributions, tests of perturbative QCD predictions for the quark polarization at
large $x$, a better understanding of quark-hadron duality, and more precise values
for higher-twist matrix elements in the framework of the Operator Product Expansion.
\end{abstract}

\keywords{Spin structure functions, nucleon structure}
\pacs{13.60.Hb, 13.88.+e , 14.20.Dh}

\maketitle

\section{INTRODUCTION}\label{s1}
One of the enduring goals in the field of hadron physics is a complete picture
of how the fundamental particles of the standard model, quarks and gluons, make up
the structure and the properties 
of the nucleon. Among other observables, the inclusive spin structure functions
$g_1$ and $g_2$ of the nucleon are a vital ingredient for this picture
(for a  review, see~\cite{Kuhn:2008sy}).
For a complete understanding of the parton structure of the nucleon, we need 
precise and comprehensive data 
not only for the proton, but also for the neutron. Since the two nucleons are isospin
partners, one can infer (assuming approximate isospin symmetry) the relative
contribution from up and down valence quarks as a function of momentum fraction $x$
 from measurements on
protons and neutrons. 
Furthermore,  fundamental sum rules concerning the difference 
between proton and neutron structure functions at all values of 
squared four-momentum transfer $Q^2$  can
be tested experimentally. The isoscalar sum of proton and neutron spin
structure functions
in the Deep Inelastic Scattering (DIS) region
 is particularly sensitive, via perturbative QCD evolution 
equations~\cite{Dokshitzer,GribovLipatov,Altarelli}, to
the gluon helicity distribution inside a longitudinally polarized nucleon. 
Moments of structure functions from proton and neutron access different matrix
elements of local operators within the Operator Product Expansion
approach~\cite{OPE1,OPE2,SSFope}. 
Finally, a better understanding of the phenomenon of
quark-hadron duality~\cite{Bloom:1970xb,Melnitchouk:2005zr} 
requires detailed studies of polarized as well as
unpolarized structure functions of both nucleons in the resonance and DIS regions.
While suitable free neutron targets do not exist, one can extract
spin structure functions for a bound neutron using polarized nuclei
like $^2$H and $^3$He, using some prescription to account for
Fermi-motion and the effective polarization of nucleons in nuclei. The results will be further affected to some
extent by Final State Interaction (FSI) effects that are presently unknown.
They have been estimated to be small in the DIS region~\cite{Cosyn:2013uoa} but may
be larger in some part of the kinematic region covered by the data reported here. In the following,
we quote results for the bound neutron without correcting for such FSI effects.
 

The CLAS (CEBAF Large Acceptance Spectrometer) collaboration at Jefferson Lab has collected
 a comprehensive set of
spin structure function data on the proton
as well as the deuteron
 over a wide range in
$Q^2 \approx 0.05 - 5$ GeV$^2$, and over a wide range of final state masses $W = 1 - 3$ GeV.
A comparable data set 
has been collected for the
neutron, using polarized $^3$He as an effective neutron target and the spectrometers
in Jefferson Lab's Hall A~\cite{Amarian:2003jy,Zheng:2004ce,Solvignon:2008hk}. However, 
nuclear binding effects have to be accounted for in a model-dependent way
in order to extract
neutron structure functions from nuclear data. In particular, in the resonance region where
cross sections and asymmetries may vary rapidly with $W$, Fermi smearing makes the
extraction of neutron results challenging and somewhat ambiguous. For those reasons,
neutron data extracted using an independent method and a different target, namely
deuterium, are highly desirable, both to check systematic uncertainties and to 
more directly access the isoscalar combination $g_1^p + g_1^n$ and its moments.
Some deuteron data in the resonance region exist from the RSS experiment~\cite{Wesselmann:2006mw},
albeit over a relatively narrow range in $Q^2$. 
Many other experiments~\cite{Abe:1998wq,Adeva:1998vw,Anthony:1999rm,Airapetian:2006vy,Alexakhin:2006vx} 
have measured spin structure functions of the deuteron in the deep inelastic
(DIS) region, $W > 2$ GeV and $Q^2 > 1$ GeV$^2$, or at small $x$~\cite{Ageev:2007du}. 
Very recently, the CLAS collaboration has published precise results from
the EG1-DVCS run on the proton and the deuteron at the highest $Q^2$ accessible with 
Jefferson Lab so far~\cite{Prok:2014ltt}.

With the experiment presented here (dubbed ``EG1b'') we collected 
a comprehensive data set on  deuteron  ($^{15}$ND$_3$) targets with nearly equal
 statistical precision and kinematic coverage as on
polarized protons ($^{15}$NH$_3$). 
The proton results will be published separately~\cite{EG1b_pfin}.
In this paper, we present our final
results  for   the asymmetry
$A_1(W,Q^2)$ and the spin structure function $g_1(x,Q^2)$ and its moments for the deuteron.
The data were obtained  in Jefferson Lab's Hall B during the time
period 2000 -- 2001. 
Previously, a  much smaller data set on the deuteron was
collected with CLAS in 1998~\cite{Yun:2002td}.
The present data set was taken with beam energies of 1.6, 2.5, 4.2 and 5.7 GeV.
Preliminary results from the highest and lowest beam energies have been 
published~\cite{Dharmawardane:2006zd,Bosted:2007aa,Prok:2008ev}.
The present paper includes, for the first time, the full data set collected with CLAS
in 2000-2001 on the deuteron, including experimental and analysis details.
We also provide,
for the first time, our  results for the corresponding (bound) neutron
structure functions, based on a somewhat
model-dependent deconvolution procedure which accounts for 
Fermi motion in the deuteron~\cite{Kahn:2008nq}.

Our analysis of the deuteron data follows closely that
for the proton data taken at the same time.
Insofar as both analyses share the same ingredients and methods, only a brief summary
is given here -- the details will be provided in the companion proton paper~\cite{EG1b_pfin}. 
However,
where the two analyses differ, we give all  details specific to the deuteron in what follows.
After a brief summary of formalism and theoretical background
(Section II), we describe
 the experimental setup (Section III) and the analysis procedures (Section IV). 
 We present the results for all measured and derived quantities, as well as 
 models and comparison to theory, in Section V, and offer our
 conclusions in Section VI.

\section{THEORETICAL BACKGROUND}\label{s2}
\subsection{Formalism}
\label{formalism}

We define the usual kinematic quantities in inclusive lepton scattering:
Incident ($E$) and scattered ($E'$) lepton energy in the lab, scattering
angle $\theta$, energy transfer $\nu = E - E'$ and squared four-momentum transfer
\begin{equation}
\label{Q2:eq}
Q^2 = -q^2 = \vec{q}^{\, 2} - \nu^2 = 4EE'\sin^2\frac{\theta}{2} .
\end{equation}
The invariant final state mass is
\begin{equation}
\label{W:eq}
W =\sqrt{M^2 + 2M\nu - Q^2} ,
\end{equation}
and the Bjorken scaling variable
\begin{equation}
\label{x:eq}
x = \frac{Q^2}{2M\nu}
\end{equation}
 in which $M$ is the nucleon mass.
The following variables are also used:
\begin{equation}
\label{gamma:eq}
\gamma = \frac{2Mx}{\sqrt{Q^2}} = \frac{\sqrt{Q^2}}{\nu} , \tau = \frac{\nu^2}{Q^2} = \frac{1}{\gamma^2} ,
\end{equation}
and the virtual photon polarization ratio
\begin{equation}
\label{epsilon:eq}
\epsilon  =  \left( 1 + 2[1 +
  \tau]\tan^2\frac{\theta}{2}\right)^{-1} .
\end{equation}

\subsection{Cross sections and asymmetries}
\label{SFsAsyms}

The observable measured in EG1b is the double spin asymmetry
\begin{equation}
\label{Aparintro:eq}
A_{||}(\nu,Q^2,E) = \frac{d\sigma^{\uparrow\Downarrow} - d\sigma^{\uparrow\Uparrow}}{d\sigma^{\uparrow\Downarrow} + 
d\sigma^{\uparrow\Uparrow}} 
\end{equation}
for inclusive
electron deuteron scattering
with beam and target spin parallel ($\uparrow\Uparrow$) or antiparallel ($\uparrow\Downarrow$) along
the beam direction.
It depends on the four structure functions
$F_1^d, F_2^d, g_1^d$ and 
$g_2^d$~\footnote{In principle, the tensor structure function $b_1$ also enters 
in the denominator, since any realistic polarized target will have a non-zero tensor polarization $P_{zz}$. 
However, in our case this is a sub-percent
 correction since  $P_{zz}$ is expected to be less than 0.1 for our target~\cite{Meyer:1985dta} and the tensor asymmetry $A_{zz}$ was measured by HERMES to be of order
 $0.01 - 0.02$~\cite{Airapetian:2005cb}.} .
Introducing the ratio $R$ of the longitudinal to transverse virtual photon absorption cross sections,
\begin{equation}
\label{Rsig}
R = \frac{\sigma_L}{\sigma_T} = 
\frac{F_2}{2 x F_1} (1+\gamma^2) - 1 ,
\end{equation}
and the variables
\begin{equation}
\label{eta:eq}
D = \frac{1 - E'\epsilon/E}{1+\epsilon R} \,\,\, \mathrm{ and } \,\,\, \eta = \frac{\epsilon\sqrt{Q^2}}{E - E'\epsilon},
\end{equation}
we can express $A_{||}$ as:
\begin{equation}
\label{g2solve:eq}
\frac{A_{||}}{D} =
(1+\eta\gamma) \frac{g_1}{F_1} +
[\gamma(\eta-\gamma)] \frac{g_2}{F_1} .
\end{equation}

Alternatively, the double spin asymmetry $A_{||}$  can also be interpreted
in terms of the two virtual photon asymmetries $A_1$ and $A_2$:
\begin{equation}
\label{Apar:eq}
A_{||} = D[A_1(\nu,Q^2) + \eta A_2(\nu,Q^2)] .
\end{equation}
Because of the relative size of the kinematic factors in Eqs.~\ref{g2solve:eq}--\ref{Apar:eq}, 
our data are mostly sensitive to $g_1$ or $A_1$, which are the main quantities of interest (see
Sections~\ref{photonabsorption} and \ref{g1}).
Given a model or other information for $F_1$, $R$ and $A_2$, $A_1$ can be
 extracted directly from Eq.~\ref{Apar:eq} and $g_1$ from
 \begin{equation}
 \label{g1fromApar}
 g_1 = \frac{\tau}{1+\tau} \left(\frac{A_{||}}{D} + (\gamma - \eta) A_2 \right) F_1 .
 \end{equation}
 Our deuteron data are not sensitive enough to $A_2$ or $g_2$ to constrain these quantities;
 instead a model based on other existing data is used (see Section~\ref{models}).

\subsection{Virtual photon absorption asymmetry}
\label{photonabsorption}

The asymmetry $A_1$ can be interpreted  in terms of transition amplitudes to specific final states (at  $W < 2$ GeV, {\em i.e.} in the resonance region) 
or in terms of the underlying quark helicity distributions (at larger $W$ and $Q^2$). 
In the former case, the measured asymmetry $A_1$  at a given value of $W$ 
gives information on the helicity structure of the combined resonant and non-resonant contributions to the
inclusive cross section, which can help to constrain the spin-isospin structure of nucleon
resonances.

In the DIS region,  $A_1(x)$ can yield information on the polarization of the valence quarks at
sufficiently large $x$  ($x \ge 0.5$), where they dominate. In the naive parton model, without taking nuclear effects into account,
the limit of $A_{1d}(x)$ at large $x$ is given as 
\begin{equation}
A_{1d} \approx  \frac{\Delta u_v + \Delta d_v}{u_v + d_v} = \frac{\Delta u_v/u_v + (d_v/u_v) \Delta d_v/d_v}{1 + d_v/u_v} ,
\end{equation}
where $u_v,d_v$ are the unpolarized up and down 
valence quark distributions and $\Delta u_v, \Delta d_v$ are the corresponding
helicity distributions.
In a $SU(6)$-symmetric, non-relativistic quark model~\cite{Greenberg:1900zza}, $\Delta u/u = 2/3$ and $\Delta d/d = -1/3$, and
$d/u = 1/2$, yielding $A_{1d} = 1/3$.
On the other hand, more advanced quark
models predict that $A_{1d}(x) \rightarrow 1$ as $x \rightarrow 1$ due to SU(6) symmetry 
breaking~\cite{isgur:prl78}.
However, even relativistic constituent quark models~\cite{Isgur:1998yb}  
predict a much slower rise towards $A_1 = 1$ than 
perturbative QCD calculations~\cite{Brodsky:1994kg,Farrar:1975yb} incorporating helicity conservation. Recently,
modifications of the pQCD picture to include orbital angular momentum~\cite{Avakian:2007xa}
have yielded an intermediate approach towards $x = 1$.
Precise measurements of $A_1$ at large $x$ and in the DIS region are therefore 
required for protons, deuterons and neutrons to establish the validity of these predictions.


\subsection{The spin structure function $g_1$}
\label{g1}
The structure function $g_1(x,Q^2)$ contains important information on the internal spin
structure of the nucleon. In the DIS limit (large $Q^2$ and $\nu$), it encodes the polarized Parton
Distribution Functions (PDFs)
 $\Delta q(x) = {q\uparrow(x)} - {q\downarrow(x)}$ for quarks with helicity
aligned {\em vs.} antialigned
with the overall (longitudinal) nucleon spin. Its logarithmic $Q^2$ dependence contains, via the
QCD evolution equations~\cite{GribovLipatov, Altarelli, Dokshitzer}, information on the analogous
helicity-dependent gluon PDFs $\Delta G(x)$ as well. The deuteron, as an approximate isoscalar nucleon target,
is particularly sensitive to $\Delta G(x)$, given a sufficiently large range
in $Q^2$. Jefferson Lab data, like those presented in this paper, can serve as a valuable
anchor point at the lowest possible $Q^2$ for NLO fits to extract $\Delta q(x)$ and $\Delta G(x)$.

In the region of lower $Q^2$, additional
scaling violations occur due to higher-twist contributions, leading to correction terms proportional to 
powers of $1/Q^2$. These corrections can be extracted 
from our data
since they cover seamlessly the transition from $Q^2 \ll 1$ GeV$^2$ to the scaling region
$Q^2 > 1$ GeV$^2$. 
In the kinematic region where $\nu$ is also small and therefore $W < 2$ GeV, the structure of $g_1$
is dominated by the contributions from nucleon resonances (similarly to $A_1$).

However, as already observed by Bloom and Gilman~\cite{Bloom:1970xb} for the unpolarized
proton structure function $F_2$, there seems to be
some duality between structure functions in the resonance region (averaged over a suitable range
in $W$) and their extrapolated DIS values at the same quark momentum fraction
$x$ or $\xi =   \frac{ |\vec{q}| - \nu}{M}$. This correspondence should be tested for both
nucleon species and for polarized as well as unpolarized structure functions to elucidate
the underlying dynamics. EG1b data have uniquely suitable kinematic coverage stretching from the
resonance to the DIS region to test whether duality holds for $g_1$. (An initial study of duality
based on part of the EG1b data has been published~\cite{Bosted:2007aa}.)

\subsection{Quasi-elastic scattering}
\label{elas}
The virtual photon asymmetries $A_1$ and $A_2$ are also defined for elastic scattering off the nucleon
and the same relationship Eq.~\ref{Apar:eq} applies. One can show that
$A_1 = 1$ in this case, and
\begin{equation}
A_2(Q^2) = \sqrt{R} = \frac{G_E(Q^2)}{\sqrt{\tau} G_M(Q^2)},
\end{equation}
where $G_E$ and $G_M$ are the electric and magnetic Sachs form factors of the nucleon.

One can also extend the definition of $g_1(x)$ and $g_2(x)$ for the nucleon to include elastic scattering,
$x = 1$:
\ba
\label{elasticgs}
g_1^{el}(x,Q^2) & = & \frac{1}{2} \frac{G_E G_M + \tau G_M^2}{1 + \tau} \delta(x-1) \nonumber \\ 
g_2^{el}(x,Q^2) & = & \frac{\tau}{2} \frac{G_E G_M - G_M^2}{1 + \tau} \delta(x-1) .
\ea

For a bound system like deuterium, one has to consider the initial state (Fermi-) motion of the struck
nucleons. In quasi-elastic inclusive scattering,
$W \lesssim  1$ GeV, both the neutron and the proton contribute (weighed by their elastic
cross sections). Alternatively, if one detects the struck proton in addition to the scattered electron
with small missing four-momentum, the 
asymmetry $A_{||}$ will be close to that on a free proton~\cite{jeschonnek09}. 
In both cases, the theoretical asymmetry can be
calculated with reasonable precision (given a realistic deuteron wave function) and therefore the
measured asymmetry can be used to extract the product of target and beam polarization (see below).

\subsection{Moments}
\label{moments}

In addition to the structure function $g_1(x)$ itself, its moments (integrals over $x$ weighted by powers of $x$) 
are of great interest. Within the Operator Product Expansion formalism, these moments can be related
to local operators \cite{OPE1,OPE2}. They are constrained by several sum rules and can be calculated directly within lattice QCD
or in effective field theories like
Chiral Perturbation Theory ($\chi$PT)~\cite{Ji:1999pd,Bernard2013aa}. 
Determining these moments over a range of $Q^2$ allows us
to study the transition from hadronic degrees of freedom at large distances (small $Q^2$) to partonic ones at small distances in our description
of the nucleon, and to extract higher twist matrix elements  that are sensitive to quark-gluon 
correlations in the nucleon.

The  first moment of $g_1$,
\begin{equation}
\label{Gamma1def}
\Gamma_{1}(Q^{2})\equiv\int_{0}^{1}g_{1}(x,Q^{2})dx ,
\end{equation}
can be related to the contribution $\Delta \Sigma$ of the quark helicities 
 to the nucleon spin in the limit of very high $Q^2$.
In particular, for the average of proton and neutron (the isoscalar nucleon approximated by the deuteron)
one has 
\begin{equation}
\label{Gamma1HiQ}
\frac{\Gamma_{1}^{p+n}(Q^2 \rightarrow \infty) }{ 2} \approx \Gamma_{1}^{d} =
\frac{5}{36} \left( \Delta u + \Delta d \right) + \frac{1}{18} \Delta s .
\end{equation}

Forming the difference between proton and the neutron yields the famous
Bjorken sum rule~\cite{Bjorken:1966jh, Bjorken:1969mm}:
\begin{equation}
\label{BjHiQ}
\Gamma_{1}^{p} - \Gamma_{1}^{n} = \frac{1}{6} a_3 = 0.211
\end{equation}
where  $a_3 = g_A = 1.267 \pm 0.004$ is the neutron axial beta decay constant.

At high but finite $Q^2$, these moments receive logarithmic pQCD corrections.
At the more modest $Q^2$ of our data, additional corrections due to higher twist 
matrix elements and proportional to powers of $1/Q^2$ become important:
\begin{equation}
\label{twistexpansion}
\Gamma_{1}(Q^{2}) = \mu_2(Q^2) + \frac{M^{2}}{9 Q^2}\left[a_{2}(Q^2)+4d_{2}(Q^{2})+4f_{2}(Q^{2})\right] \cdots
\end{equation}
Here, $\mu_2$ is the leading twist contribution given by Eq.~\ref{Gamma1HiQ} plus pQCD corrections,
$a_2$ and $d_2$ are due to target mass corrections and $f_2$ is a twist-4 matrix element that
contains information on quark-gluon correlations and has been calculated using
quark models~\cite{PhysRevD.73.074016}, QCD sum rules~\cite{Balitsky:1989jb} 
and other approaches like lattice QCD~\cite{Dolgov:1998js}.

In addition to the leading first moment, odd-numbered higher moments of $g_1$ can be defined as
$\Gamma_1^n = \int_0^1 dx x^{n-1} g_1(x), \, n = 3,5,7,...$. These moments are dominated by high $x$
(valence quarks) and are thus particularly well determined by data in Jefferson Lab kinematics. They
can also be related to hadronic matrix elements of local operators
or evaluated with Lattice QCD methods. The third moment $\Gamma_1^3$ is related to the
matrix element $a_2$ above.

In the limit of very small photon virtualities $Q^2$, moments of spin structure functions
can be connected to observables in Compton scattering. In particular,
the first moment is constrained by the Gerasimov-Drell-Hearn (GDH) sum 
rule~\cite{Gerasimov:1965et, Drell:1966jv} in the limit $Q^2 \rightarrow 0$:
\be
\label{Gamma1slope}
\left. \frac{d \Gamma_1(Q^2)}{d Q^2} \right|_{Q^2 = 0} = - \frac{\kappa^2}{8 M^2} ,
\ee
where $\kappa$ is the anomalous magnetic moment of the  nucleon. Higher order
derivatives at the photon point are, in principle, calculable via 
$\chi$PT~\cite{Ji:1999pd,Bernard2013aa}.
Therefore, measuring $\Gamma_1$ over the whole range in $Q^2$ yields a stringent
test of our understanding of strongly interacting matter at all length scales.

Extending the analysis of low-energy Compton amplitudes to higher orders, 
one can get additional generalized sum 
rules~\cite{Gorchtein:2004jd}. In
particular, one can generalize the forward spin polarizability, $\gamma_0$,
to virtual photons:
\be
\label{gamma0}
\gamma_0(Q^2) = \frac{16 \alpha M^2}{Q^6} 
\int_0^{1} x^2 \, \big[g_1(x,Q^2) - \gamma^2 g_2(x,Q^2)\big]\,dx .
\ee
Once again, this generalized spin polarizability can be calculated  using
$\chi$PT~\cite{Bernard2013aa}.

\subsection{From nucleons to the deuteron}
\label{deuteron}
Most of the previous discussion is focused on the interpretation of spin structure functions of the nucleon
(proton and neutron). Where appropriate, we indicate how this interpretation may be modified when 
the nucleons are embedded in deuterium. Here, we want to discuss in more detail how
the nuclear structure of the deuteron affects the measured asymmetries and structure functions.

In the most simple-minded picture, all observables on the deuteron can be considered (cross section weighted) averages
of the corresponding proton and neutron observables. Spin observables are further modified by the fact that
even in a fully polarized deuteron, the nucleon spins are not 100\% aligned due to the D-state component
of the wave function. To first order, this can be corrected by applying a reduction factor 
$(1- 1.5 P_D)$ to all nucleon spin observables inside deuterium~\cite{CiofidegliAtti:1996uc},
 with $P_D \approx 4-6\%$ being the 
D-state probability (according to the results from recent nucleon-nucleon potentials~\cite{Argonne}). Taking this factor into account, the spin structure functions $g_1^d(x)$ and $g_2^d(x)$ 
of the deuteron are reasonably well approximated by the average of the proton and neutron ones, as long
as $x$ is not too large ($x < 0.6$) and $W$ is not in the resonance region ({\it i.e.}, 
$W > 2$ GeV). Moments of these
structure functions can be considered as relatively ``safe'' since the integration averages over effects like Fermi 
motion~\cite{CiofidegliAtti:1996uc}.

In the valence region of moderate to large $x$ and in the resonance region, Fermi-smearing due to the 
intrinsic motion of the nucleons inside deuterium as well as nuclear binding and FSI become more important, because
structure functions vary rapidly in this region with $W$ or $x$. These binding effects can be partially
modeled by convoluting the free nucleon structure functions with the momentum distribution of nucleons
inside deuterium. In our analysis, we use a recent convolution model by 
Melnitchouk et al.~\cite{Kahn:2008nq,Malace:2009dg}
that properly treats the effects of finite momentum transfer $Q^2$. 

On the other hand, no universal model of the effects of FSI over the whole kinematic region
covered by our data is available; we therefore do not correct for those effects. Similarly,  potential off-shell effects (due to the negative binding energy of nucleons inside deuterium), including perhaps a modification of the nucleon structure (the EMC effect) and non-nucleonic degrees of freedom (mesons~\cite{PhysRevC.51.52},
 $\Delta \Delta$ components~\cite{Frankfurt:1981mk,Frankfurt:1988nt}
  and perhaps more exotic quark structures~\cite{Schmidt:2000pj}) may play a role. Since no universally accepted model for these effects exists, we
present our results with the caveat that they are for bound neutrons only.
Given the small binding energy (-2.2 MeV) and large average inter-nucleon distance
(of order 4 fm) in deuterium, we expect these effects to be significantly smaller than in more tightly bound
nuclei. However, a comparison with neutron spin observables obtained from measurements on
$^3$He can be a valuable check on the size of nuclear binding corrections. Ultimately, the best approach to extracting
free neutron information would be to apply the method of spectator tagging (pioneered for unpolarized
structure functions  in the recent ``BONuS'' 
experiment~\cite{Tkachenko:2014aa} at Jefferson Lab).

\section{THE EXPERIMENT}\label{s3}
The EG1b experiment  took place at Jefferson Lab over  a seven month period in 2000-2001. It used the
highly polarized (up to 85\%) electron beam produced by the  Continuous Wave  
Electron Beam Accelerator (CEBAF), with energies from 1.6 GeV to nearly 6 GeV and currents
of 0.3 to 10 nA in the experimental Hall B. Detailed descriptions of the accelerator
and its strained GaAs polarized electron source can be found in  
Refs.~\cite{CEBAF:ref,Sinclair:2007ez,Kazimi:2004zv,Stutzman:2007ny}.

The beam polarization was intermittently monitored using a M\o ller polarimeter, and the beam
position and intensity distributions were measured with a set of beam monitors. The  amount
of beam charge delivered to the Hall for a given time interval was measured with a Faraday cup (FC).
The signal from this  FC was recorded separately for each beam polarization and
gated by the data acquisition live time.
In order to avoid local heating and depolarization,
the beam was rastered over the face of the target in a spiral pattern, using two magnets upstream 
from the target. 

The target consisted of cells containing samples of polarized hydrogen ($^{15}$NH$_3$), 
deuterium  ($^{15}$ND$_3$), carbon, or no solid material (``empty target'') that could be alternatively inserted in the beam.
These cells were suspended in a
liquid $^4$He bath at about 1 K. The target material was polarized inside a 5 T solenoidal field along the beam
axis,
using the method of dynamic nuclear polarization (DNP) described in~\cite{DNP:ref,orientation:ref, Crabb:1997cy}. 
The target polarization was monitored by an NMR system. Typical values of about 30\% deuteron polarization
along or opposite to the beam direction were achieved during the experiment.

Scattered electrons (and other particles) were detected with the CEBAF Large Acceptance Spectrometer (CLAS) \cite{Mecking:2003zu}  in Hall B. CLAS employs a toroidal magnetic field and several layers of detectors
in six identical sectors surrounding the beam axis for an acceptance of nearly $2 \pi$ in azimuth. Electrons
were detected in the scattering angle range from $8^\circ$ to about $50^\circ$. Three regions of 
drift chambers (DC)~\cite{Mestayer:2000we} determine charged particle trajectories, followed by 
Cherenkov counters (CC)~\cite{Adams:2001kk} and electromagnetic calorimeters 
(EC)~\cite{Amarian:2001zs} for electron identification,
while timing is provided by a scintillation
counter (SC) system~\cite{Smith:1999ii}. For EG1b, the trigger was optimized for inclusive electrons
and required a coincidence between signals above threshold in the EC and the CC.

The experimental setup and operation
will be described in detail in the upcoming companion paper on our proton results~\cite{EG1b_pfin}.

\section{DATA ANALYSIS}\label{s4}
\subsection{Data set}
Data on the deuteron (ND$_3$) were taken with seven different beam energies and two opposite polarities of the CLAS torus magnetic field. For positive (+) polarity, electrons are bent towards the beam line, and
for the negative (-) polarity, away from it. 
The in-bending (+) configuration gives access to the largest
scattering angles and allows CLAS to run with its highest possible luminosity of
$L = 2\cdot10^{34}$ cm$^{-2}$s$^{-1}$. Therefore, we used this configuration to collect the highest
$Q^2$ points for each beam energy. In the out-bending (-) configuration, electrons were
detected down to the smallest accessible 
scattering angle of $8^\circ$, extending the data set to lower $Q^2$.

\begin{figure}
\centering
\includegraphics[width=8.0cm]{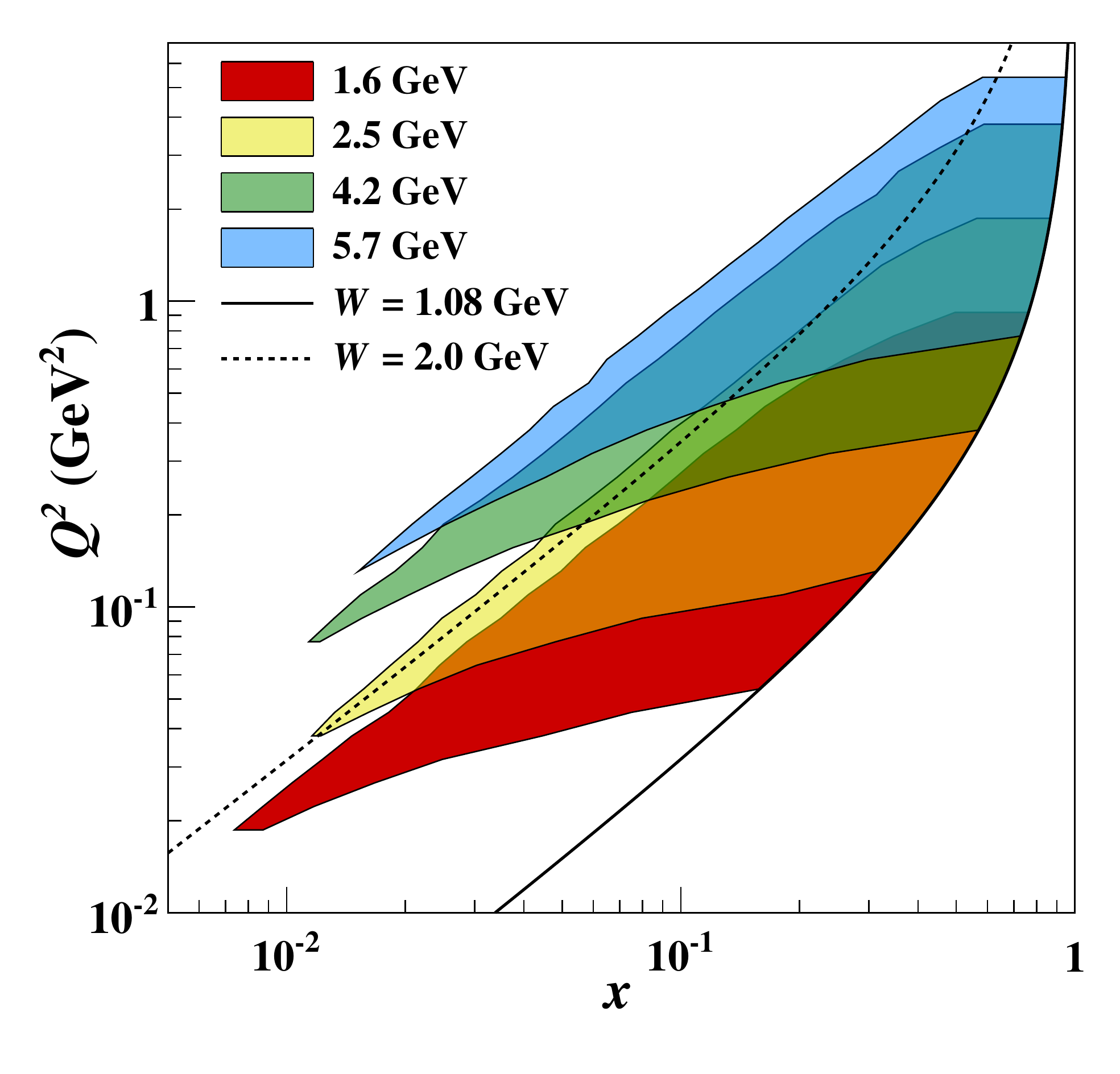}
\caption[Kinematic coverage of EG1b]{(Color Online) Kinematic coverage in $Q^2$
  vs. $x$ for each of the 4 main electron 
 beam energy groupings used in the EG1b experiment. The solid and dotted lines
denote the $W = 1.08$ and $W = 2.00$ GeV thresholds, respectively. The coverage for
proton (NH$_3$) and deuterium (ND$_3$) targets was nearly identical.}
\label{kinrange:fig} 
\end{figure}

In all, data
were collected in 11 specific combinations ($1.606+$, $1.606-$, $1.723-$,
$2.561+$, $2.561-$, $4.238+$, $4.238-$, $5.615+$, $5.725+$,
$5.725-$, $5.743-$) of beam energy (in GeV) and main torus
polarity ($+$,$-$), hereby referred to as ``sets''. Sets with similar
beam energy comprise four groupings with nominal average energies of
1.6, 2.5, 4.2 and 5.7 GeV.
The kinematic coverage of the data for each of the 4 energy groupings  
is depicted in Fig.~\ref{kinrange:fig}.


\subsection{Data selection}
\label{cuts}

After following the standard calibration procedures for all CLAS detector elements,
the raw data were converted into a condensed data summary tape  (DST) format
containing track and particle ID
information. 
Quality checks ensured that malfunctioning detector
components, changes in the target and/or potential sources of false
asymmetries did not contaminate the data. DST files not meeting the
minimal requirements were eliminated from analysis.

Event selection criteria were applied to identify scattered electrons
and to minimize the background from other particles, primarily $\pi^-$.
These criteria, based on the signals from the CC and the EC,
 will be discussed in detail in~\cite{EG1b_pfin}.
We ascertained that the remaining $\pi^-$ contamination of our electron
sample was less than 1\% over the whole kinematic range. 
For this reason, we assign a 1\% systematic uncertainty on our extracted
asymmetries as an upper limit for any remaining pion contamination effect.

\begin{figure}
\centering
\includegraphics[width=8.5cm]{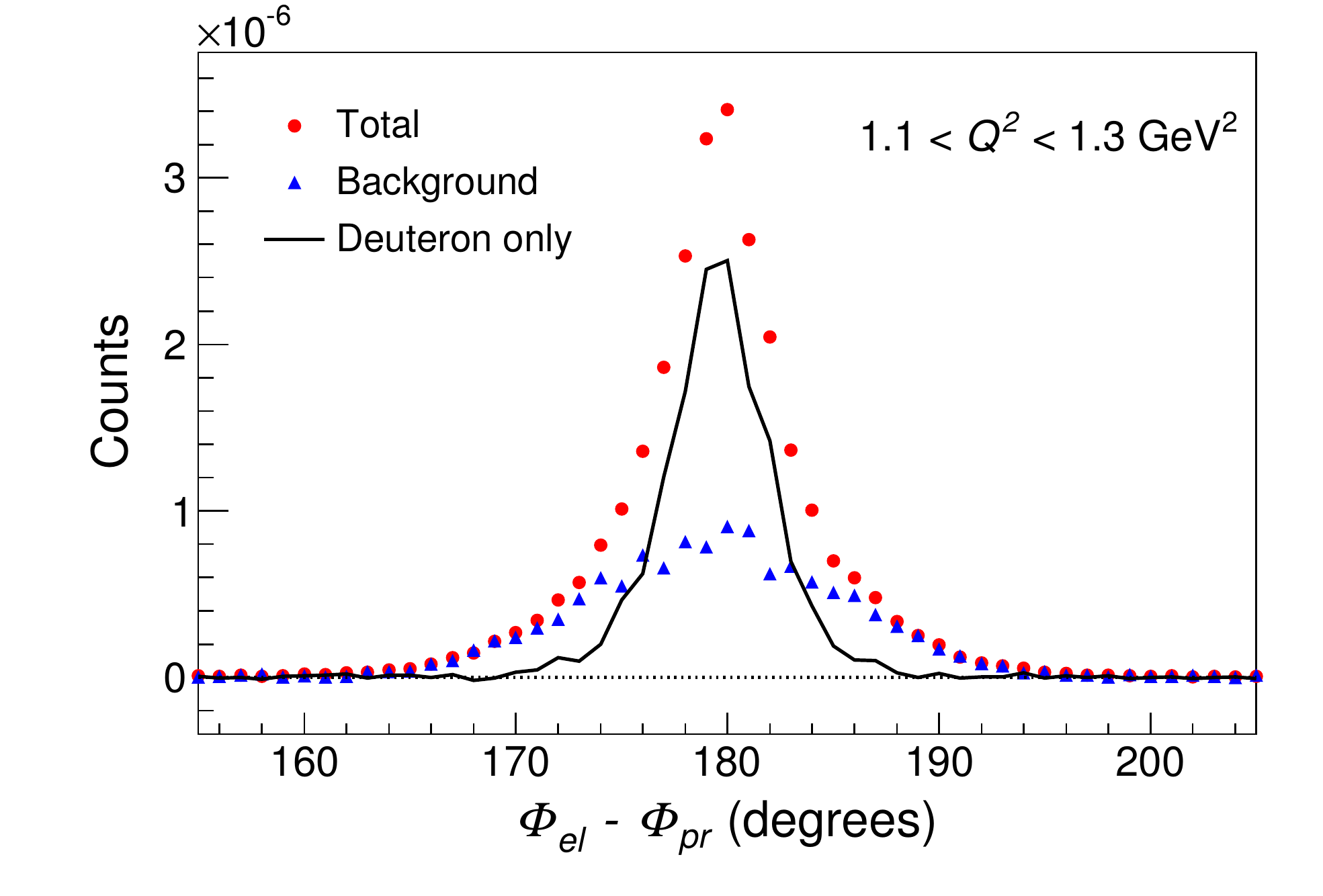}
\caption{(Color Online) Distribution of quasielastic $d(e,e'p)$ events versus the angle $\phi$ between
the azimuth of the scattered electron and the azimuth of the observed proton.
 The background due to nitrogen,
liquid $^4$He and various foils is strongly suppressed by the cuts described in the text, leading to a 
relatively clean
signal from the deuteron component (solid line)
 of the target. A final cut is applied from $\phi = 177^\circ$ to $183^\circ$.}
\label{deltaphi:fig} 
\end{figure}

For the determination of the product of beam and target polarization ($P_bP_t$, see below) as
well as kinematic corrections, we also required a sample of quasi-elastic $(e,e'p)$ events.
We selected $ep$ coincidences through a timing cut of $\pm 0.8$ ns on the difference
between the reconstructed electron and proton vertex time. Quasi-elastic events
were selected through cuts on $W$, 0.89 GeV $\le W \le$ 1.01 GeV, missing energy
(of the unobserved nuclear remnant) of $\le 0.08$ GeV (kinetic), and on the difference between
the polar ($|\Delta \theta| \le 2^\circ$) and azimuthal ($|\Delta \phi| \le 3^\circ$) angles
of the detected proton and the reconstructed direction of the virtual photon.
These cuts were optimized to include most of the $ep$ coincidences from quasi-elastic
scattering on the deuteron, while the contribution from the other target components
(nitrogen, $^4$He and foils) was much suppressed due to the wider nucleon momentum
distributions in these nuclei (see Fig.~\ref{deltaphi:fig}).

\subsection{Event corrections}

The track information for particles in the DSTs is based on an ideal detector and has
to be corrected for various effects from detector materials and imperfections. Among other
corrections, energy loss due to ionization in the target (both for the incoming and the scattered
electron), multiple scattering angle deviations (compared to the average vertex of all particles in an
event), and known deviations of the target magnetic field from the ideal version implemented
in the reconstruction software were used to correct each track within an event. 

The reconstruction software also assumes that a track originates on the nominal central axis ($x = y = 0$) of
CLAS. In reality, the beam is rastered over a circle of about 1.5 cm diameter,  
whose center 
 is typically offset
by a few mm from the nominal axis. Since the raster position can be inferred from the currents in the raster magnets, the
reconstructed vertex was corrected for this offset.

\begin{figure}[bht!]
\centering
\subfigure
{
  \includegraphics[trim=0cm 3.8cm 0cm 0cm, clip=true, width=7cm]{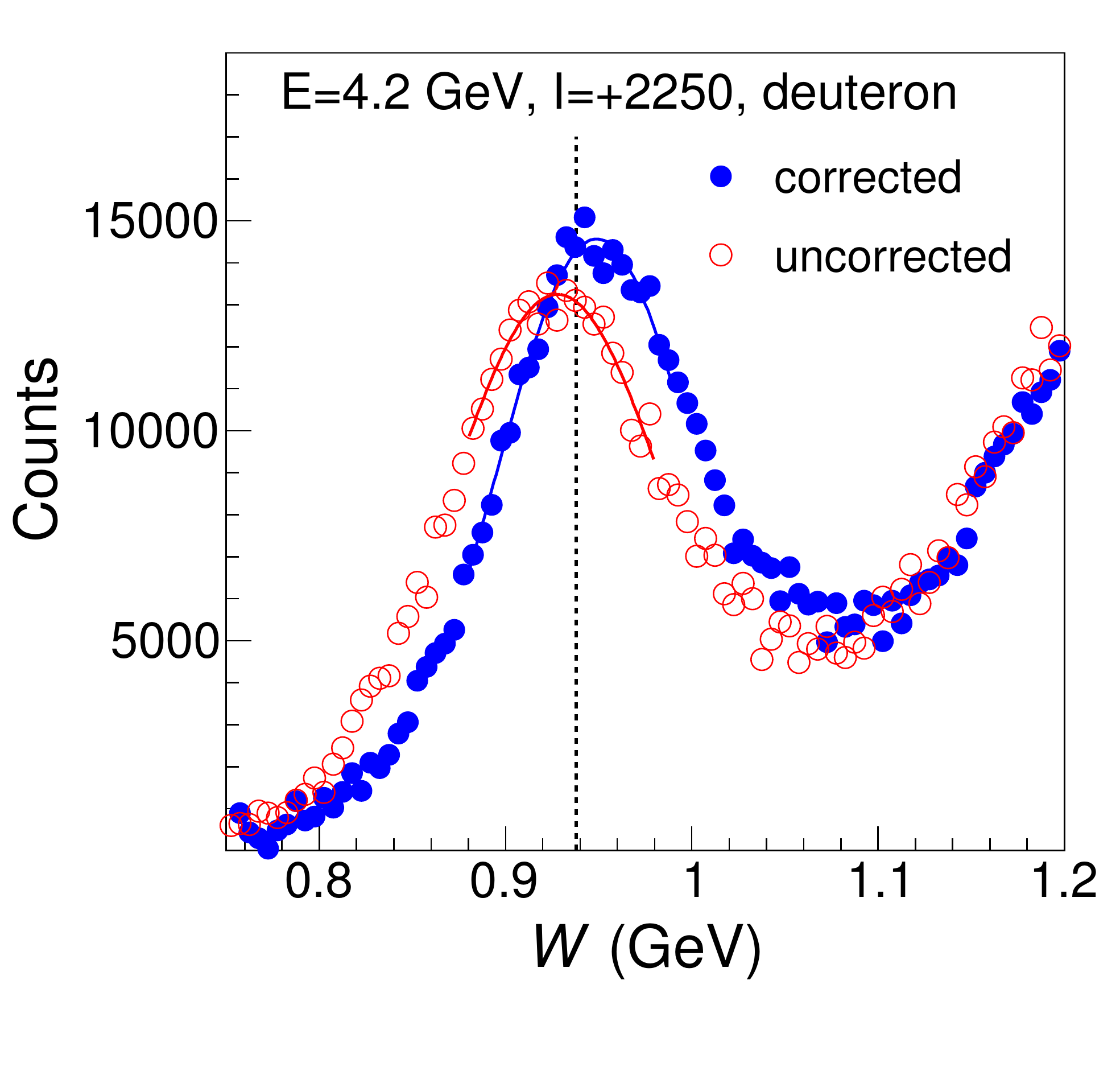} 
}
\subfigure
{
  \includegraphics[trim=0cm 0cm 0cm 0.6cm, clip=true, width=7cm]{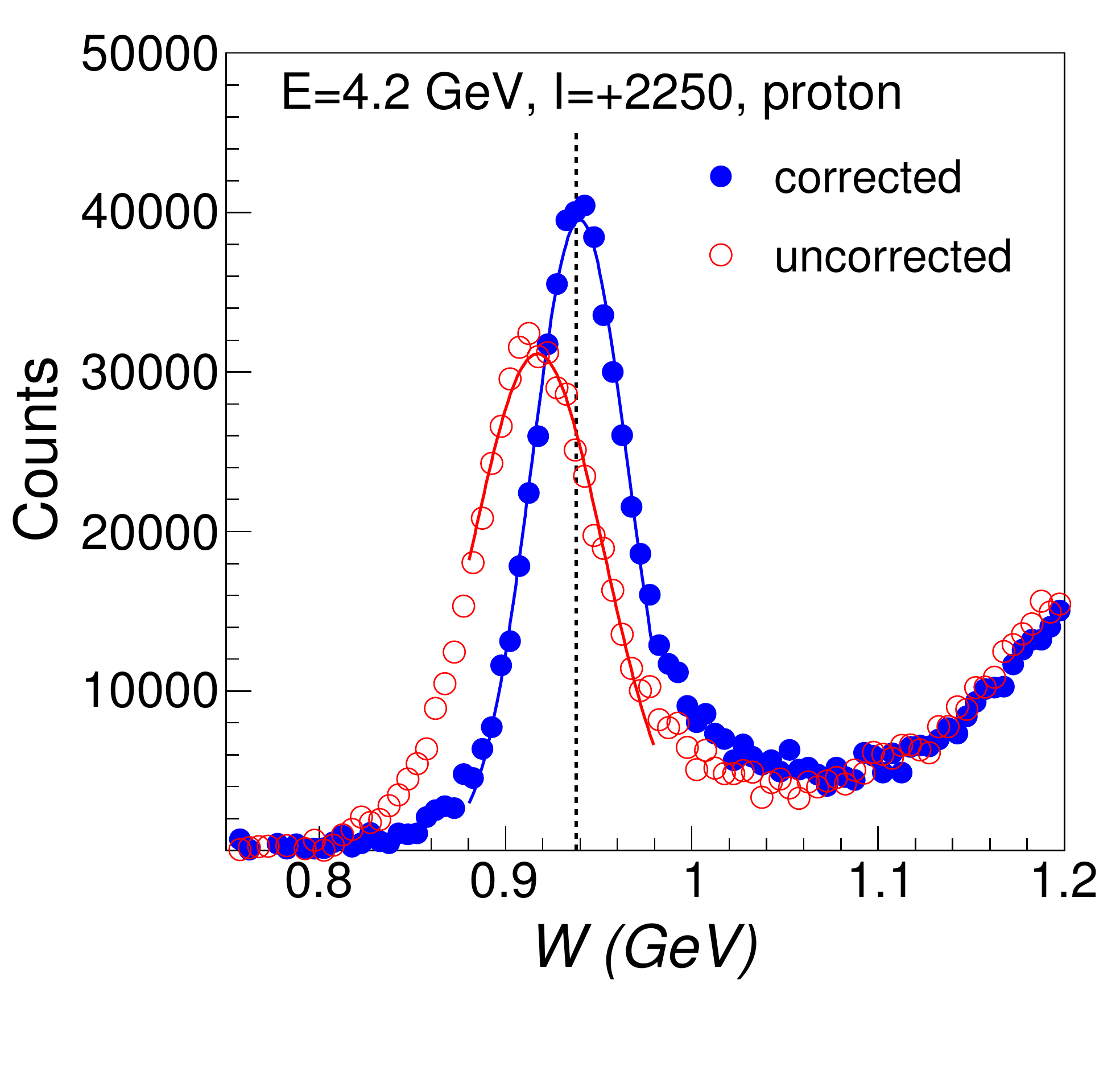} 
}
\caption[$W$ spectrum before and after the kinematic corrections for the 4.231 GeV inbending data set.]{
(Color Online) Missing mass $W$ before (red-hollow) and after (blue-solid) the kinematic
corrections for the 4.238+ data set for NH$_{3}$ (top) and ND$_{3}$ (bottom) targets. 
The corrections decreased the distribution width and centered the mean
value of the (quasi-)elastic peak on the nucleon mass.
}
\label{WPeak:fig}
\end{figure}

The position and orientation
of the drift chambers in space and the detailed three-dimensional shape of the torus magnetic field are
not known with absolute precision; an empirical parametrization of their deviations from the ideal detector
was obtained from a fit to data from the companion experiment on the proton~\cite{EG1b_pfin}.
We used four-momentum conservation in
fully exclusive events like H$(e,e^\prime p)$ and
H($e,e^\prime p \pi^+ \pi^-)$ to optimize the fit parameters. This parametrized correction for
particle momenta and scattering angles
was
then applied to each track. The resulting improvement of the resolution in the 
missing mass $W$ is shown in Fig.~\ref{WPeak:fig}.

A final correction was applied to the integrated beam charge measured by the Faraday Cup, to account for beam
loss between the target and the FC due to multiple scattering and due
to dispersion by the target magnetic field.

\subsection{From raw to physics asymmetries}

For each combination of beam energy, torus polarity and target polarization, electron tracks
were sorted by kinematic bins and were counted
separately for positive ($N^+$) and negative ($N^-$) beam helicity, where ``+'' refers to a beam helicity
antiparallel to the direction of the target polarization. These counts were normalized to the corresponding
integrated Faraday charges, $n^\pm = N^\pm / FC^\pm$. Only events coming from complete pairs of
``beam buckets'' with opposite helicity were counted to avoid false asymmetries; we also ascertained that, after
averaging over all target polarizations, the residual beam charge asymmetry
$(FC^+ - FC^-)/(FC^+ + FC^-)$ was less than $10^{-4}$. These normalized counts were used to form the
 raw asymmetry
\begin{equation}
\label{rawasym:eq}
A_{raw} = \frac{n^+-n^-}{n^++n^-}
\end{equation}
in each kinematic bin.
This raw asymmetry was then converted to the desired physics asymmetry
$A_{||}$ (Eq.~\ref{Aparintro:eq}) by applying a series of corrections which we now discuss in sequence.

\subsubsection{Dilution factor}\label{DFcorr}
The dilution factor $F_{DF}\equiv n_d/n_A$ is defined as the ratio of
events from polarizable nuclei of interest (here, deuterons bound in ammonia, $n_d$) to
those from all components of the
full ammonia target ($n_A$). It is calculated directly
from the radiated cross-sections on all components of the target.
 In terms of densities ($\rho$), 
material thicknesses ($\ell$) and cross-sections per nucleon ($\sigma$), 
\begin{equation} 
n_d \propto \frac{6}{21}\rho_A\ell_A\sigma_d
\end{equation} 
and
\begin{multline}
\label{nA:eq} 
n_A \propto \rho_{Al}\ell_{Al}\sigma_{Al} +
 \rho_{K}\ell_{K}\sigma_{K} \\+   \rho_A\ell_A(\frac{6}{21}\sigma_d +
 \frac{15}{21}\sigma_N) + \rho_{He}(L-\ell_A)\sigma_{He} ,
\end{multline}
with the subscripts $A$, $Al$, $K$, $N$, and $He$ denoting deuterated
ammonia ($^{15}$ND$_3$), aluminum foil, kapton foil, nitrogen ($^{15}$N)
and helium ($^4$He), respectively. The acceptance-dependent 
proportionality constant is identical in both of the above relations for a given kinematic bin.
Inclusive scattering data from the
empty (LHe) and $^{12}$C
targets were analyzed to determine the total target cell length ($L$)
and effective ND$_3$ thickness ($\ell_A$) using similar equations.

The required
cross-sections were calculated from a fit to world data for $F_1$ and
$F_2$ for protons and neutrons, using a Fermi-convolution model
to fit
inclusive scattering data on nuclear targets, including EG1b data from
$^{12}$C, solid $^{15}$N and empty (LHe) targets \cite{Christy:2007ve,Bosted:2007hw}.
The nuclear EMC effect was parametrized using SLAC data
\cite{Norton:2003cb}. Radiative
corrections used the treatment of Mo and Tsai \cite{Mo:1968cg}; external
Bremsstrahlung probabilities incorporated all material thicknesses in CLAS from
the target vertex through the inner layer of the DC. 

\begin{figure}
\centering
  \includegraphics[width=8cm]{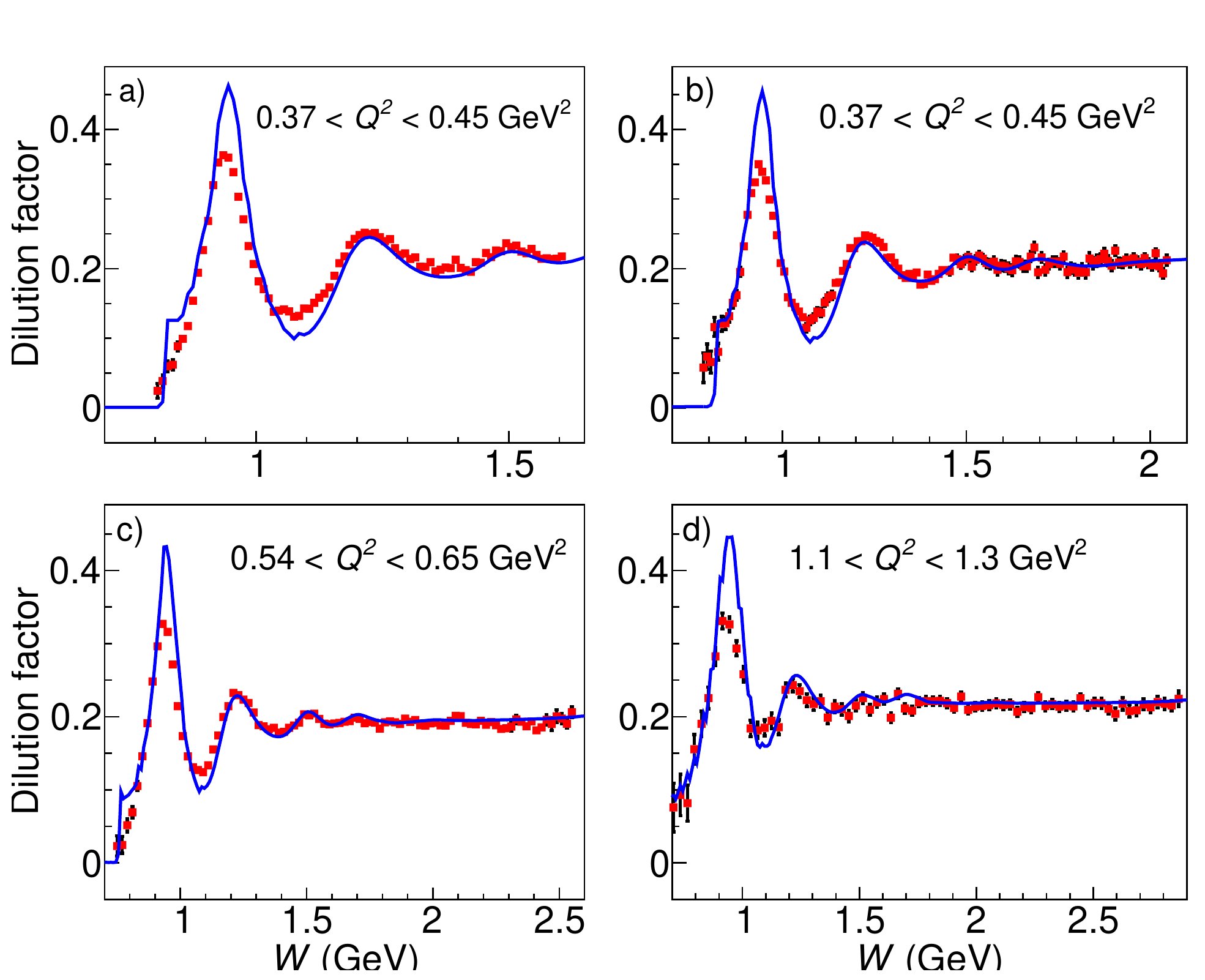}
\caption[Dilution factors plotted vs. $W$]{
(Color Online) Dilution factors as a function of $W$, shown at four different beam energies (1.6$+$ (top left), 2.5$-$ (top right), 4.2$-$ (bottom left) and 5.7$-$ (bottom right)). The results from 
our standard method (using cross section models, see text) are shown as blue lines, 
while the results from the data-based
method (see text) are shown as the red data points.
}
\label{DF_method12:fig}
\end{figure}

Dilution factors $F_{DF}$ were  calculated for each
data set and used to correct the raw asymmetry,
\begin{equation}
A_\mathrm{undil} = \frac{A_\mathrm{raw} }{F_{DF}},
\end{equation}
to get the undiluted asymmetry due to deuterons in the target.
We checked our results for $F_{DF}$ from the ``standard method'' described above
against a previously developed ``data-based
method''~\cite{Fatemi:2003yh,Dharmawardane:2006zd,Prok:2008ev} 
that uses 
a simple model of neutron/proton  cross-section ratios
 to express the background in the ammonia
target in terms of the counts from carbon and empty targets.
Values of $L$ and $\ell_A$ varied by less than 2\% between the two
methods. Figure~\ref{DF_method12:fig} shows the result from both methods for four
kinematic bins. For the inelastic data, $W > 1.1$ GeV, the dilution factors from the cross-section based
standard method were more precise and were used to correct the raw asymmetries. We used the data-based
method only in the quasi-elastic region  $W < 1.08$ GeV
 (for the determination of beam and target polarization in one case,
see below) and to subtract the background from exclusive $d(e,e'p)n$ events (see Fig.~\ref{deltaphi:fig}).
This is because finite detector resolution effects (which are not included in the cross section model) 
significantly affect the
shape of sharply peaked spectra in the quasi-elastic region, making the data-driven method 
more reliable.

The densities and thicknesses of all target materials were varied
within their known tolerances to determine systematic uncertainties; only
the variations of $\rho_C\ell_C$ and $\rho_{He}$ had any significant ($>$1\%)
effect on $F_{DF}$. Uncertainties due to the cross-section model were
estimated by the comparison of $F_{DF}$ to a third-degree polynomial
fit to the data-based dilution factors determined by the alternate method.

\subsubsection{Beam and target polarizations ($P_bP_t$)}
\label{pbpt:sec}

The second major factor to consider when extracting the physics asymmetry $A_{||}$ is the
product of beam and target polarization by which the measured asymmetry must be divided.

Because NMR measurements provided accurate target polarization
measurements only near the edge of the target cell \cite{Keith:2003ca}
(which was not uniformly exposed to the beam), we determined the
polarization product $P_bP_t$  directly from our data, using
quasi-elastic $d(e,e'p)n$ and (in one case) $d(e,e')$ events. Here,
we made use of the fact that the theoretical asymmetry in this case
depends only on the electromagnetic form factors of the proton and the neutron, see
Section~\ref{elas}, which are well-known~\cite{Arrington:2003qk}, giving us reliable predictions
of $A_{||}$. After correcting for the (relatively smaller) dilution of this asymmetry 
from non-deuterium components of the target, we can directly divide
the measured $A_{||}$ by the theoretical one to extract $P_bP_t$:
\begin{equation}
P_bP_t = \frac{A_\mathrm{meas}^\mathrm{QE}}{F_{DF}\;A_\mathrm{theo}^\mathrm{QE}} .
\end{equation}

We used the value for $P_bP_t $ obtained from inclusive quasi-elastic events only in one case, for the 
1.6 -1.7 GeV outbending configuration
runs. In that case, too few of the protons from $d(e,e'p)n$ were detected in CLAS
for a reliable determination of $P_bP_t$. We used a cut of 0.89 GeV $\le W \le$ 1.01 GeV
to define quasi-elastic events. While this method yields a smaller statistical uncertainty,
it has greater systematic uncertainty because of larger background contributions; therefore, a systematic
uncertainty of 10\% was assigned to this particular $P_bP_t$ value.

For all other configurations, we determined $P_bP_t$ using exclusive $d(e,e'p)n$ events within the cuts
listed in Section~\ref{cuts} which have very little background from nuclear target 
components (see Fig.~\ref{deltaphi:fig}). We used a detailed Monte Carlo simulation, including
Fermi motion of the proton inside the deuteron, to calculate the theoretical asymmetry.
For both methods, the nuclear background was determined using the data-driven method
mentioned in Section~\ref{DFcorr}. As a cross check, we compared these results to the values 
derived from inclusive quasi-elastic scattering, and found them generally to be consistent 
within the statistical uncertainty. 

\begin{figure}
\centering{ 
\includegraphics[width=\linewidth]{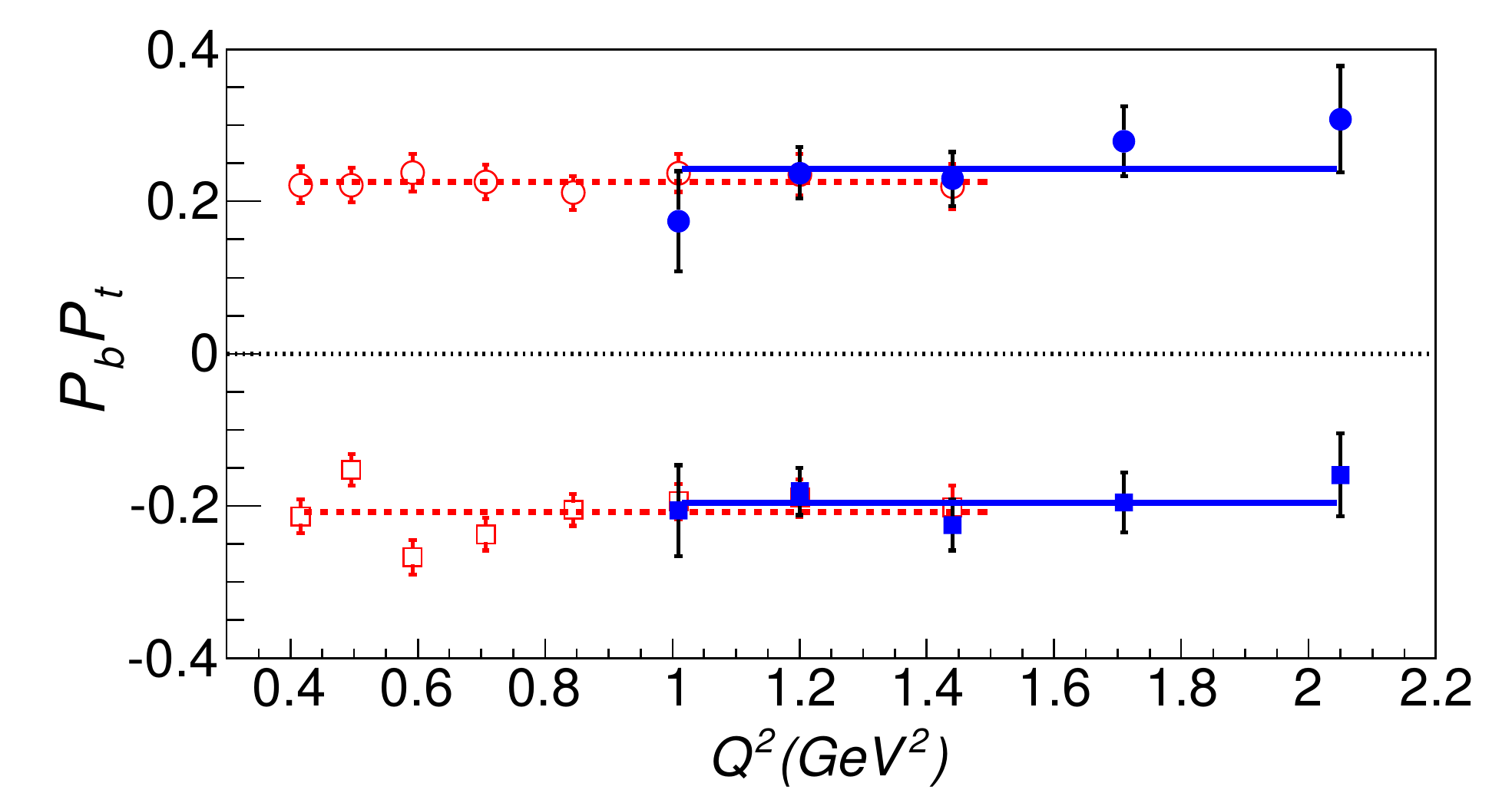}
}
\caption[$P_bP_t$ values for different data sets for ND$_{3}$ target.] 
{ 
(Color Online) $P_bP_t$ values for the 2.5 GeV inbending data sets. The plot shows the resulting $P_bP_t$ values
extracted independently for each $Q^2$ bin with available data. 
The results from the exclusive (blue filled symbols) and the inclusive (red open symbols) methods are shown. The corresponding constants fit to the data are also shown as lines: the solid blue line is for the exclusive and the dashed red line is for the inclusive method.
}
\label{pbpttargetpol:fig}
\end{figure}

The derived $P_bP_t$ values were checked for consistency across $Q^2$ for
each beam energy, torus current and target polarization direction.
 Sample $P_bP_t$ values across $Q^2$
for 2 beam energies are shown in Figure
\ref{pbpttargetpol:fig}. Across all beam energies, $P_bP_t$ values ranged from  0.1 to 0.28, with most values between
0.15 and 0.25.
We varied each of the values of $P_bP_t$ individually by the larger of one (statistical) standard deviation and the
difference between the exclusive and inclusive results to assess the systematic uncertainty of all physics quantities due to $P_bP_t$.

\subsubsection{Polarized nitrogen and target contamination corrections}
Apart from the  dilution of the measured asymmetry by nucleons embedded in  nitrogen,
helium and other target materials (Section~\ref{DFcorr}), there are 
additional small modifications of this asymmetry due to polarized target nucleons outside of deuterium.

First, it is well-known that the $^{15}$N nuclei in the ammonia molecules become 
somewhat  polarized as well.
 Equal Spin Temperature (EST) theory predicts the  polarization
ratio between two spin-interacting nuclear species in a homogenous medium as the ratio
of their magnetic moments: $P_{^{15}N}/P_{^2H} \approx
\mu_{^{15}N}/\mu_{^2H}$. However, experimentally, it was found that the $^{15}$N polarization is 
somewhat smaller than that~\cite{Crabb:1995xi}.
Using a simple shell model 
description~\cite{RondonAramayo:1999da}
of the $^{15}$N nucleus, this polarization is carried by a single proton in the $1p_{1/2}$ shell, which means
that this proton is spin-polarized to -33\% of the nucleus. The measured magnetic moment of
$^{15}$N suggests a somewhat smaller spin polarization, so that the overall contribution from nitrogen
to the measured asymmetry can be approximated by that of a  bound proton with
polarization  $P_p^{bound}$ between 8\% and 16\% of the
deuteron polarization. Accordingly, we subtracted a correction of
$1/3 \times P_p^{bound} \times A_p \sigma_p^{bound} / \sigma_{d} \approx (0.026 \pm 0.014) A_p$ from the measured
asymmetry, where the factor $1/3$ accounts for the
 three deuteron nuclei per nitrogen nucleus in ammonia.

A second contamination to the measured asymmetry comes from isotopic impurities of the deuterated
ammonia, with some deuterons replaced by protons. Typical contaminations quoted in the 
literature~\cite{Abe:1998wq}
are around 1.5\%. We did a careful study~\cite{mayer2012}
 that showed a $^1$H contamination of up to about 3.5\% during EG1 (which was included in the dilution factor); 
 however, according to this study at most one-half
of these extra protons were polarized (the remainder are presumably bound
 in molecules like H$_2$O and are unpolarized). The degree of polarization of these protons can be
 estimated as  $P_p/P_d \approx 1.2 - 1.5$, again according to EST and empirical evidence~\cite{RondonAramayo:1999da}.
 The net effect is an additional term proportional to $A_p$ that has to be subtracted from the measured
 asymmetry. The total correction for bound and free polarized protons in the target is between
 $0.027A_p$ and $ 0.051 A_p$. We took the median of this range to correct our data (using a model of the
 asymmetry $A_p$ based on our proton results~\cite{EG1b_pfin}) and $1/2$ of its spread to estimate systematic
 uncertainties. 
 An additional correction due to the very small contribution
 of $^{14}$N nuclei (less than 2\% of our ammonia sample) was too small to be applied but was
 included in the overall systematic uncertainty.

Quasi-elastic $d(e,e'p)n$ events are also affected by the various target contaminations discussed above.
We applied a corresponding correction to our extraction of $P_bP_t$ (Section~\ref{pbpt:sec}).

\subsubsection{Other background subtractions}
Dalitz decay of neutral pions \cite{Dalitz:1951aj} and Bethe-Heitler
processes~\cite{Gehrmann:1997qh} 
can produce $e^+e^-$ pairs at or near the vertex, contaminating the
inclusive $e^-$ spectrum. This contamination was at most a few percent of the data rate (at high $W$)
and was measured by comparing positron and electron rates for runs with opposite torus polarity.
We also measured the positron asymmetry and found it consistent with zero.
We subtracted this pair-symmetric
 background using the measured rate and assuming zero asymmetry. To estimate
the corresponding systematic uncertainty, we instead applied a correction assuming a
constant positron
asymmetry within the range of
values we measured. We also used the change in
 the correction after varying the rate within its uncertainty
as a second contribution to the overall systematic uncertainty for this background. 


\begin{figure}[htb!]
\centering
 \includegraphics[trim=0cm 1.89cm 0cm 0.4cm,clip=true, width=8cm]{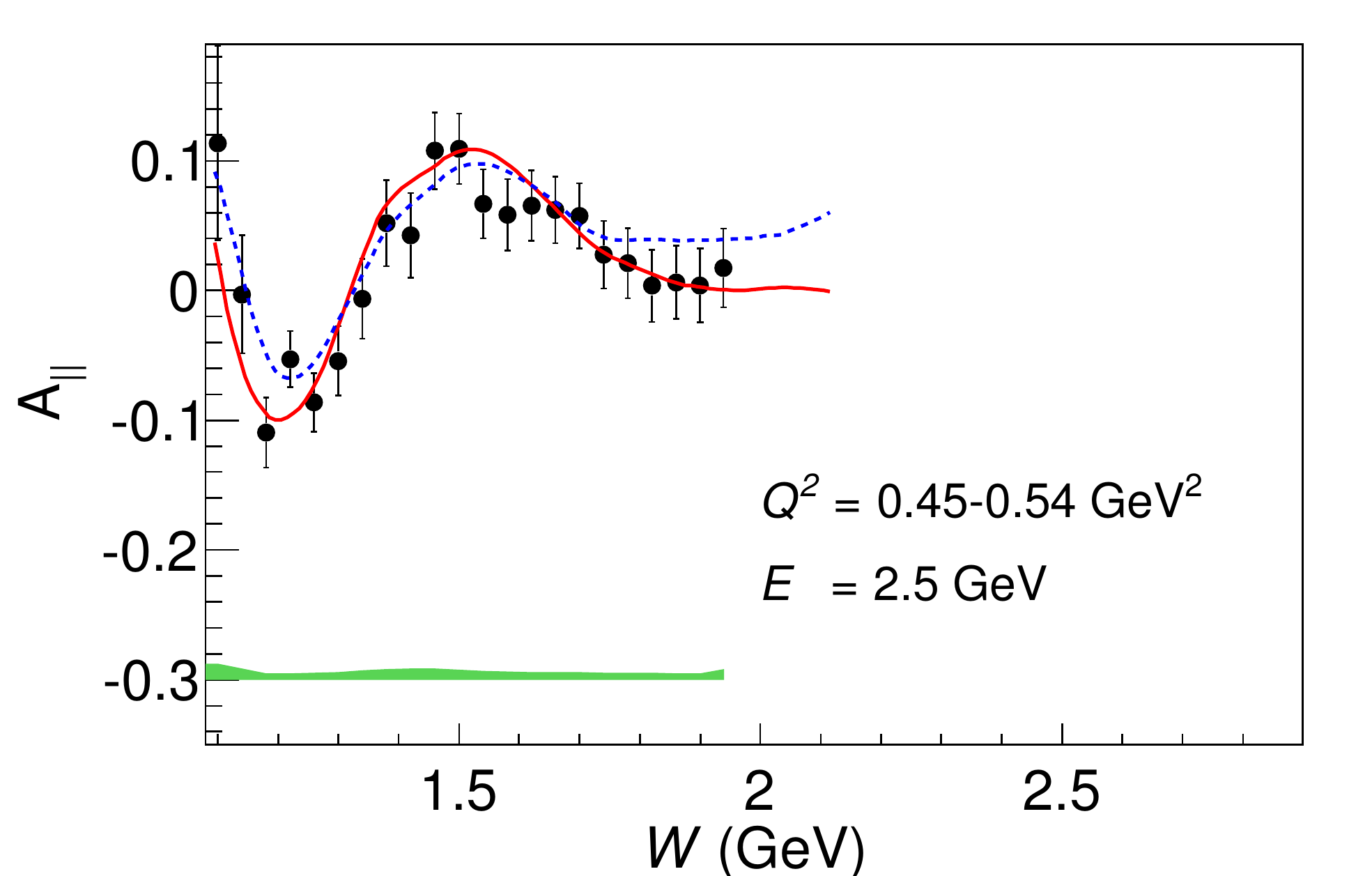}
 \includegraphics[trim=0cm 1.89cm 0cm 0.4cm,clip=true, width=8cm]{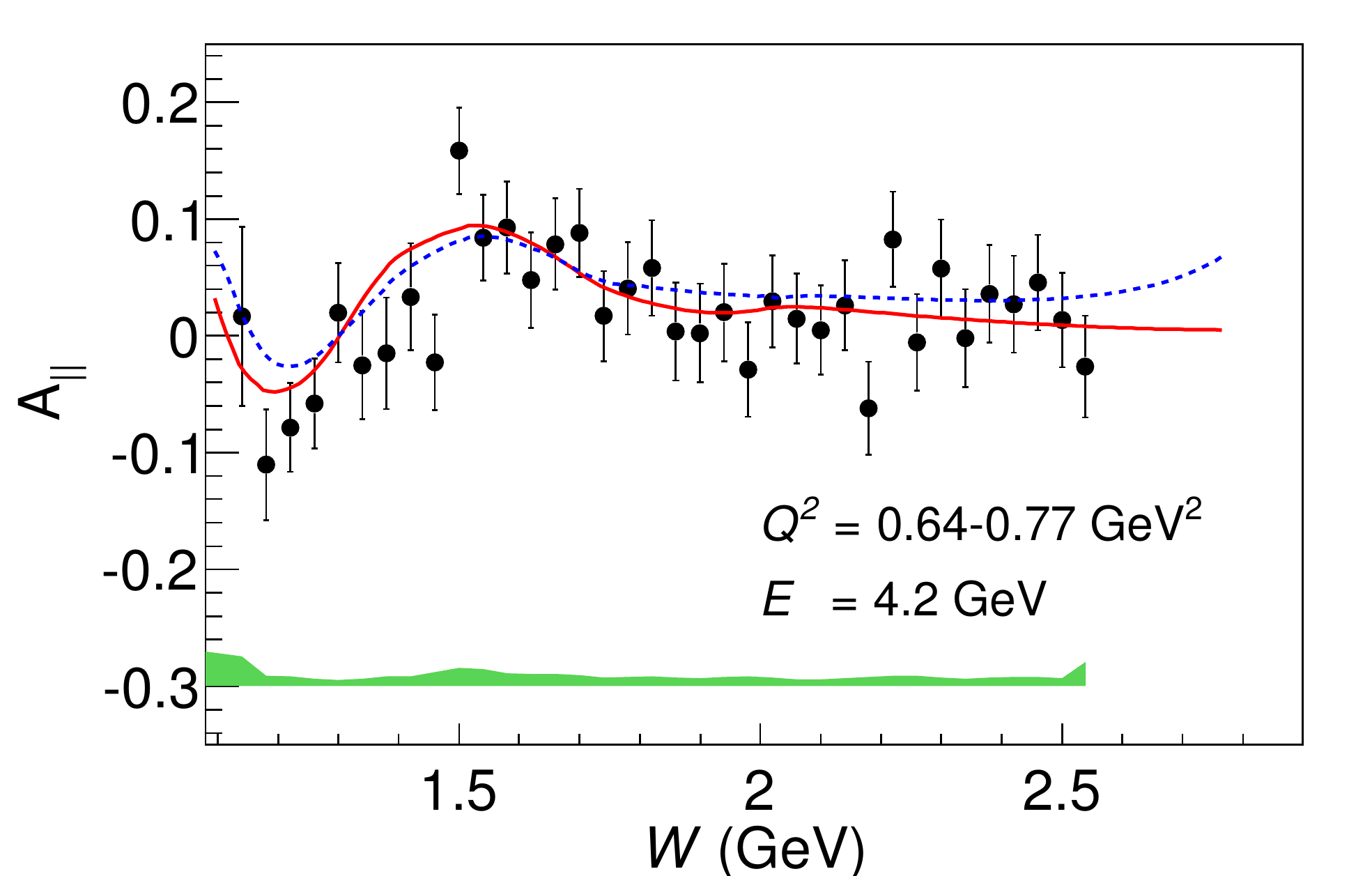}
 \includegraphics[trim=0cm 0cm    0cm 0.4cm,clip=true, width=8cm]{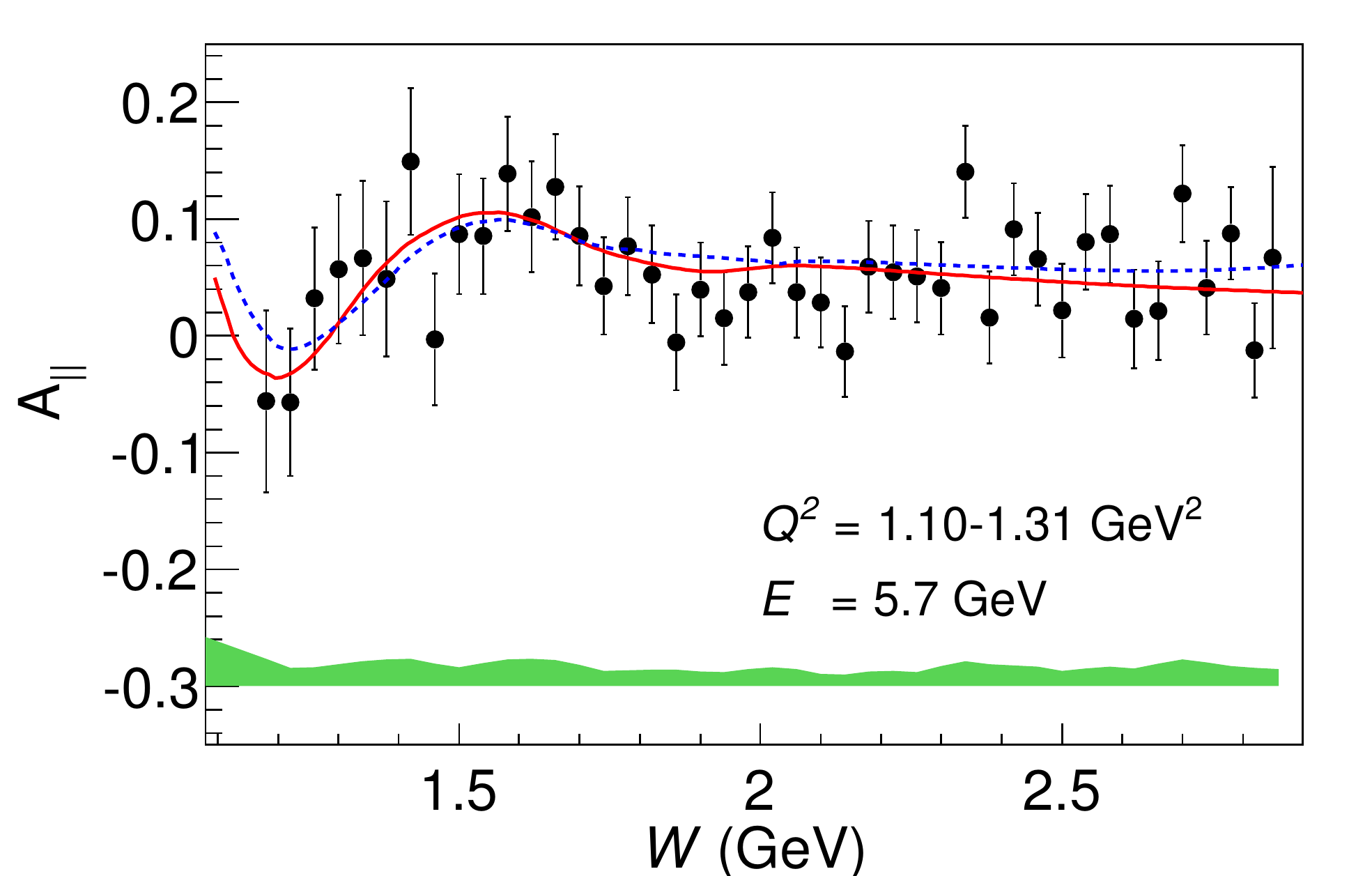}
\caption [$A_{||}$ versus $W$ shown together with the radiative corrections.]
{(Color Online) Representative results for the fully corrected double-spin asymmetry
$A_{||}$ versus final state invariant mass $W$ for three different $Q^2$ bins and beam energies.
The red-solid line represents our model parametrization of $A_{||}$ (see Section~\ref{models}).
The dashed blue lines represent the model including radiative effects. The difference between those
lines corresponds to the magnitude of radiative corrections applied. The error bars reflect statistical
uncertainties while the shaded bands at the bottom of each plot represent the total systematic uncertainties.
}
\label{Apar_ver1:fig1}
\end{figure}

\subsubsection{Radiative corrections}
\label{radcorr:sec}
Radiative corrections to the measured asymmetries $A_{||}$ were computed using the program 
RCSLACPOL, which was developed at SLAC for the spin structure function experiment 
E143~\cite{Norton:2003cb}.
Polarization-dependent internal and external corrections were calculated according to the prescriptions
in Ref.\ \cite{Kukhto:1983pv} and Ref.\ \cite{Mo:1968cg}, respectively.  

We compared the calculated double spin asymmetry   with radiative effects turned on, $A_r$, to
the Born asymmetry, $A_B$, calculated with the same models (see Section~\ref{models}). We determined
parameters $f_{RC}$ and $A_{RC}$ for each kinematic bin, allowing us to write the Born asymmetry as
\begin{equation}
A_B = \frac{A_r}{f_{RC}} + A_{RC} ,
\end{equation}
where $f_{RC}$ is a radiative dilution factor accounting for the count rate fraction 
from the elastic and quasi-elastic
tail within a given bin. This correction was then applied to all data.
Figure~\ref{Apar_ver1:fig1} shows a few examples for the magnitude of the correction, together with the
final data for the Born asymmetry $A_{||}$.

Systematic uncertainties on these corrections were estimated by running RCSLACPOL for a range of reasonable variations of the models
for $F_2$, $R$, $A_1$ and $A_2$ (see Section~\ref{models})
 and for different target thicknesses and cell lengths,
$\ell_A$ and $L$.  The changes due to each variation were added in quadrature and the
square root of the sum was taken as the systematic uncertainty on radiative effects.


\subsubsection{Systematic uncertainties}
\label{syserr:sec}
Estimation of systematic uncertainties 
on each of the observables discussed in the following
section was done by varying a particular input parameter, model
or analysis method, rerunning the analysis, and recording the
difference in output for each of the final asymmetries, structure
functions and their moments. 
Final systematic uncertainties attributable to each altered
quantity were then added in quadrature to estimate the total
uncertainty.
Note that for each quantity of interest 
($A_1, g_1, \Gamma_1$) the systematic uncertainty was calculated by this
same method (instead of propagating it from other quantities), 
therefore ensuring that all correlations in these sources were properly
taken into account.


\begin{table}[htbp] \centering
\vspace{0.5cm}
\begin{tabular}{|c|c|}
\hline
Systematic uncertainty & Typical range (in \% of $g_1/F_1$)\\
\hline
\hline
Pion and $e^+e^-$ contamination & 0.0\% -- 1.0\%\\
\hline
Dilution Factor&  1.8\% --  2.7\%  \\
\hline
Radiative corrections &3.5\% -- 5.7\% \\
\hline
$P_bP_t$ uncertainty & 6\% -- 22\% \\
\hline
Model uncertainties & 2.0\% -- 5.0\% \\
\hline
Polarized Background & 1.0\% -- 1.7\%  \\
\hline
Total &10\% -- 23\%  \\
\hline
\end{tabular}
\caption{Table of typical magnitudes for various systematic uncertainties.}
\label{syserr:table}
\end{table}

Most sources of systematic uncertainties have been discussed above.
These sources
include kinematic shifts, bin averaging,  
target parameters (radiative corrections), nuclear dilution model, structure function models,
$P_bP_t$ uncertainty for each individual data set, and background contaminations.
The relative magnitudes of these various contributions to the systematic uncertainty,
for the case of
the ratio $g_1/F_1$,  are listed in Table~\ref{syserr:table}. 
The results shown in 
the next section incorporate these systematic uncertainties.

\section{RESULTS AND COMPARISON TO THEORY}\label{s5}
\subsection{Results for $A_1 + \eta A_2$}
{\label{sec:A1A2}}

\begin{figure}[htb!]
\centering
  \includegraphics[trim=0cm 2.15cm 0cm 0cm,clip=true, width=8cm]{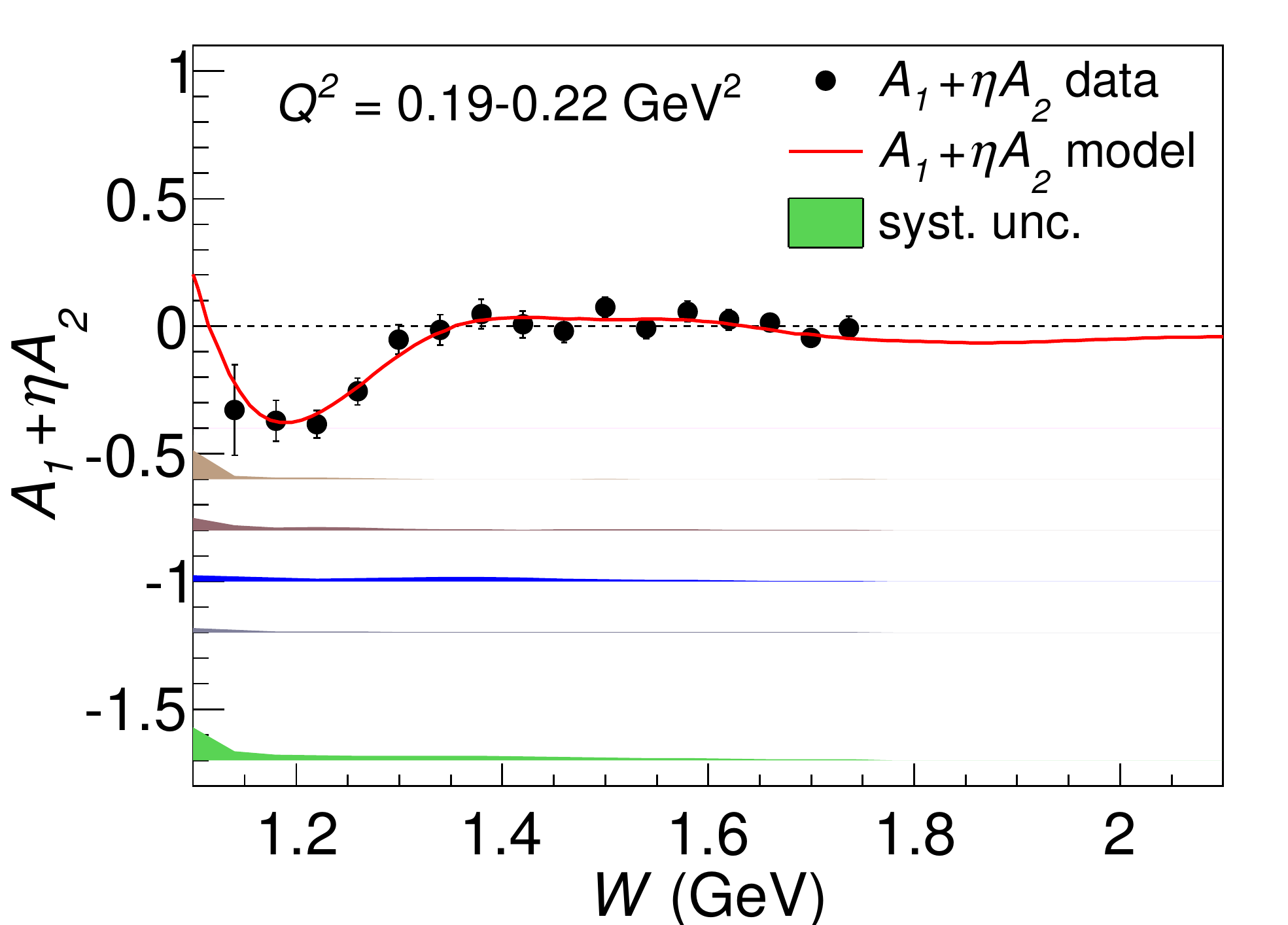} 
  \includegraphics[trim=0cm 0cm 0cm 0.4cm,clip=true, width=8cm] {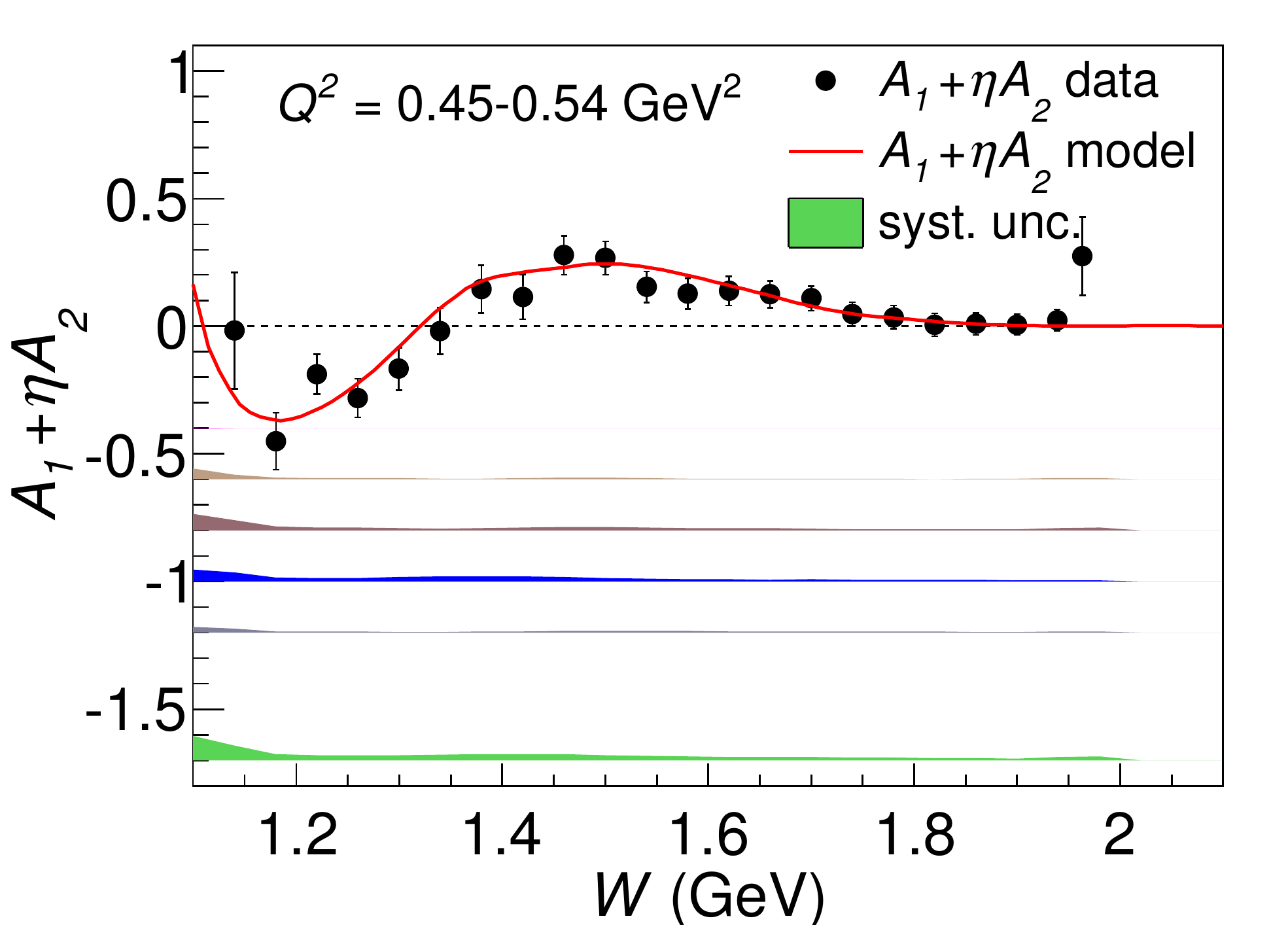} 
\caption[$A_1 + \eta A_2$ versus $W$ together with different sources of systematic uncertainties.]
{(Color Online) Representative values for the double-spin asymmetry
$A_1 + \eta A_2$ versus final state invariant mass $W$ . 
The top panel is for 0.16 GeV$^2 \leq Q^2 \leq 0.19$ GeV$^2$ (1.6 GeV data) and the bottom panel for 
0.45 GeV$^2 \leq Q^2 \leq 0.54$ GeV$^2$ (2.5 GeV data).
The red-solid line represents our model parametrization
 of $A_1 + \eta A_2$ (see Section~\ref{models}). The shaded band at the bottom (green) is the total systematic uncertainty. The individual contributions are offset  vertically, from top to bottom: pion and pair symmetric contamination (-0.4; barely visible); dilution factor (-0.6); $P_b P_t$ (-0.8); models plus radiative corrections (-1.0); and polarized background (-1.2).
}
\label{A1A2W_ver1:fig1}
\end{figure}

\begin{figure}[htb!]
\centering
  \includegraphics[trim=0cm 2.15cm 0cm 0cm,clip=true, width=8cm]{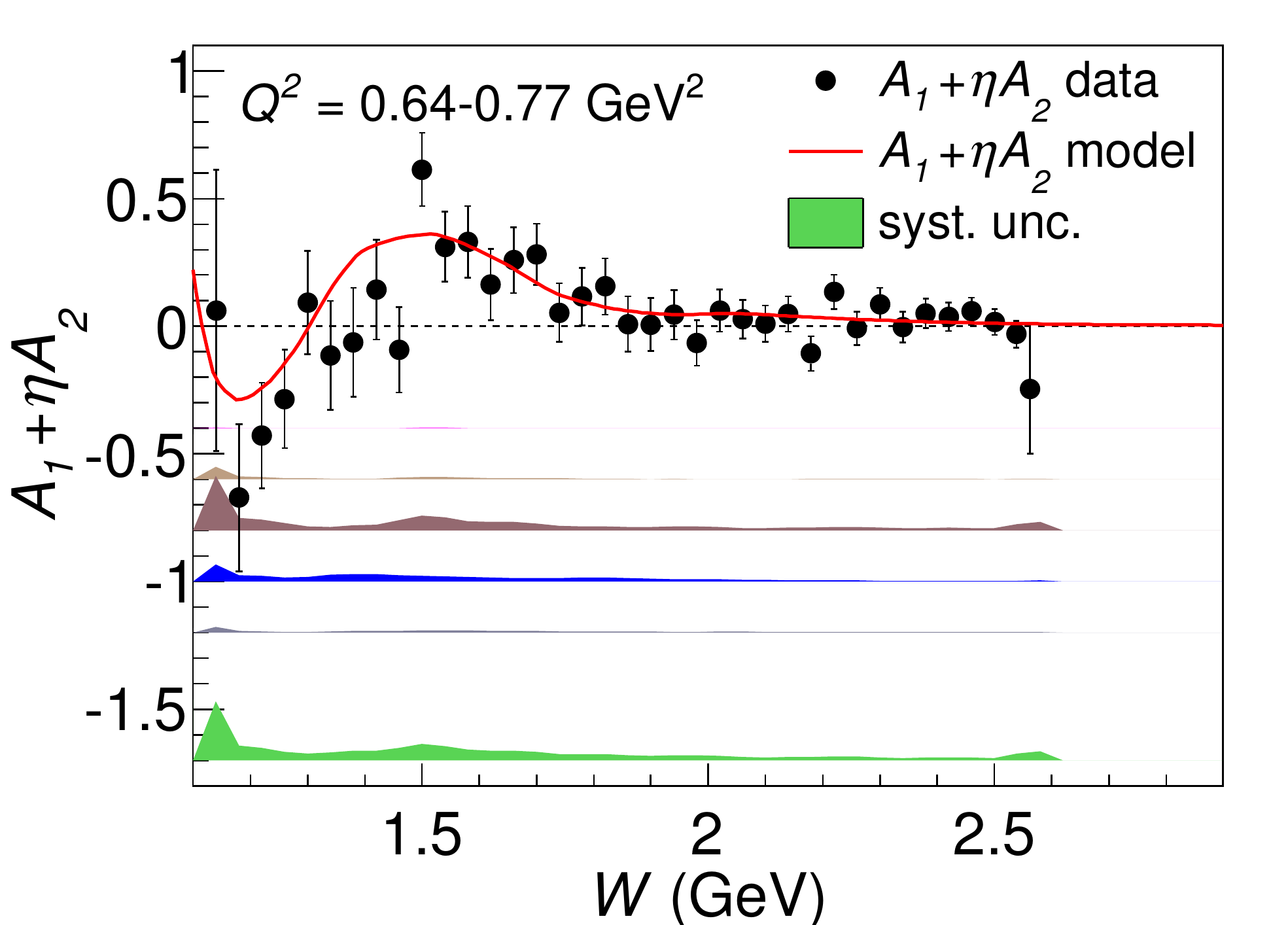} 
  \includegraphics[trim=0cm 0cm 0cm 0.4cm,clip=true, width=8cm] {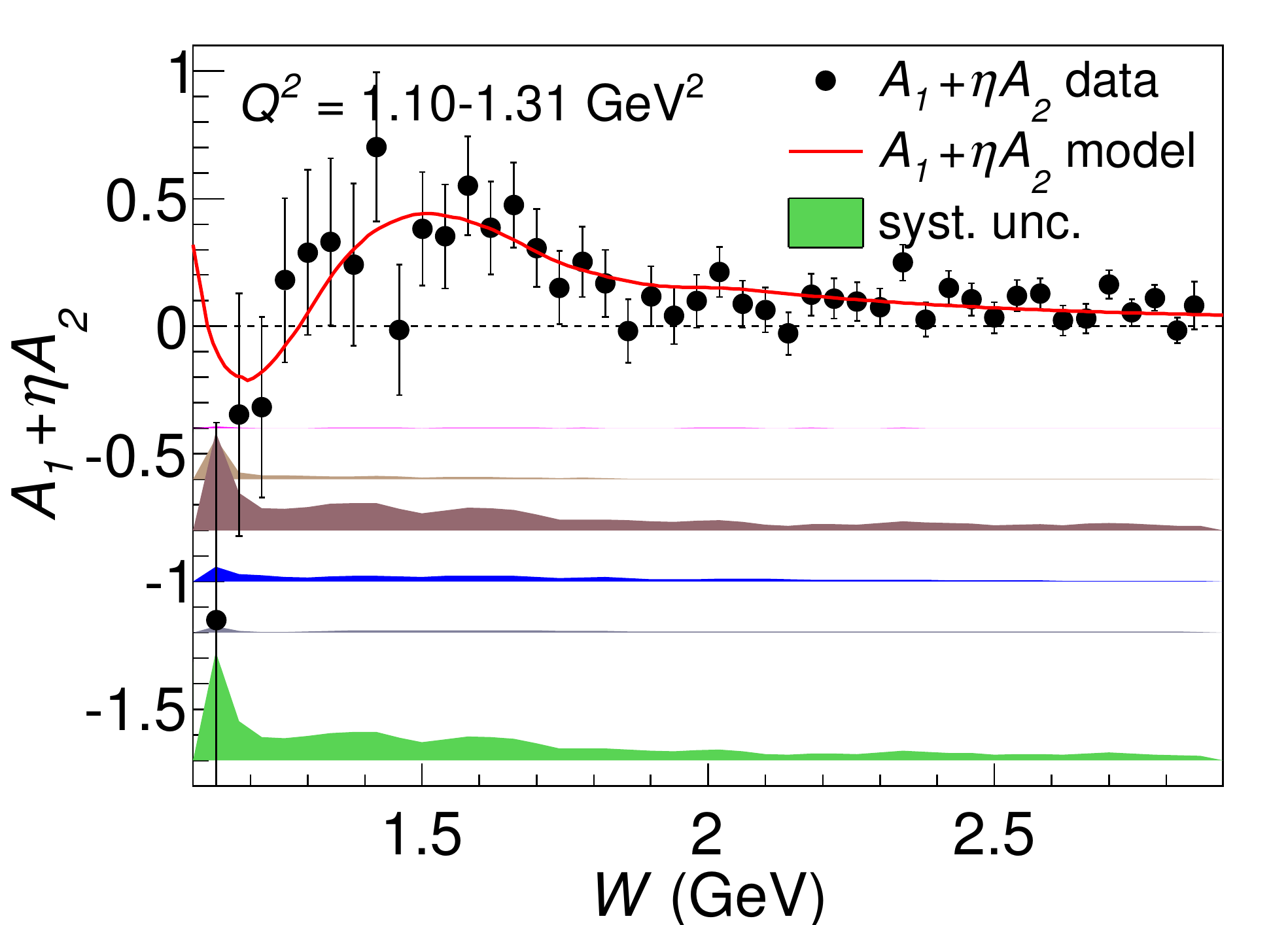} 
\caption[$A_1 + \eta A_2$ versus $W$ together with different sources of systematic uncertainty.]{
Same as Fig.~\ref{A1A2W_ver1:fig1} except for two higher beam energies.
The top panel is for 0.64 GeV$^2 \leq Q^2 \leq 0.77$ GeV$^2$ (4.2 GeV data)
 and the bottom panel for 1.1 GeV$^2 \leq Q^2 \leq 1.3$ GeV$^2$ (5.7 GeV data).}
\label{A1A2W_ver1:fig2}
\end{figure}

In this section, we present our final results for all quantities of interest:
$A_1$, $g_1$ and moments for the deuteron and the bound neutron.
 As a first step, we divide the fully corrected Born asymmetry $A_{||}$ by the depolarization factor $D$
(Eq.~\ref{eta:eq}) to extract the combination $A_1 + \eta A_2$ for each bin in $W$ and $Q^2$ and each beam
energy. Results for similar beam energies (e.g., 1.6 and 1.7 GeV) and inbending and outbending torus
polarization are combined into averaged values for four nominal energies (1.6 GeV, 2.5 GeV, 4.2 GeV and 5.7 GeV), weighted by their statistical precision. We checked that in each case, the data sets that we combined agree with each
other within statistical and systematic uncertainties.
Figures~ \ref{A1A2W_ver1:fig1} and \ref{A1A2W_ver1:fig2} show the results for $A_1 + \eta A_2$  for selected $Q^2$ bins and for each of the four standard energies. The systematic uncertainties from different contributing sources are also shown as shaded bands at the bottom of each plot. For most kinematics, the largest contribution to the systematic uncertainty 
is due to the beam and target polarization, with some contribution from the dilution factor and radiative
corrections. We note that our data for all 4 beam energies are well described by our model  (see Section~\ref{models}) as
indicated by the red solid line. 

Our results for $A_1 + \eta A_2$
have the least theoretical bias from unmeasured structure functions like $A_2$ and $F_1$,
and are therefore the preferred choice for NLO fits that will include our data in the high-$Q^2$, 
high-$W$ region, like the fit
by the JAM collaboration~\cite{JMO13}. They can be found in the CLAS 
database~\cite{DataBase} and in the Supplemental Material~\cite{SupMat} for this paper.

\subsection{The virtual photon Asymmetry $A_{1}$}
{\label{sec:A1}}

\begin{figure}[htb!]
\centering
  \includegraphics[trim=0cm 2.15cm 0cm 0cm,clip=true, width=8cm]  {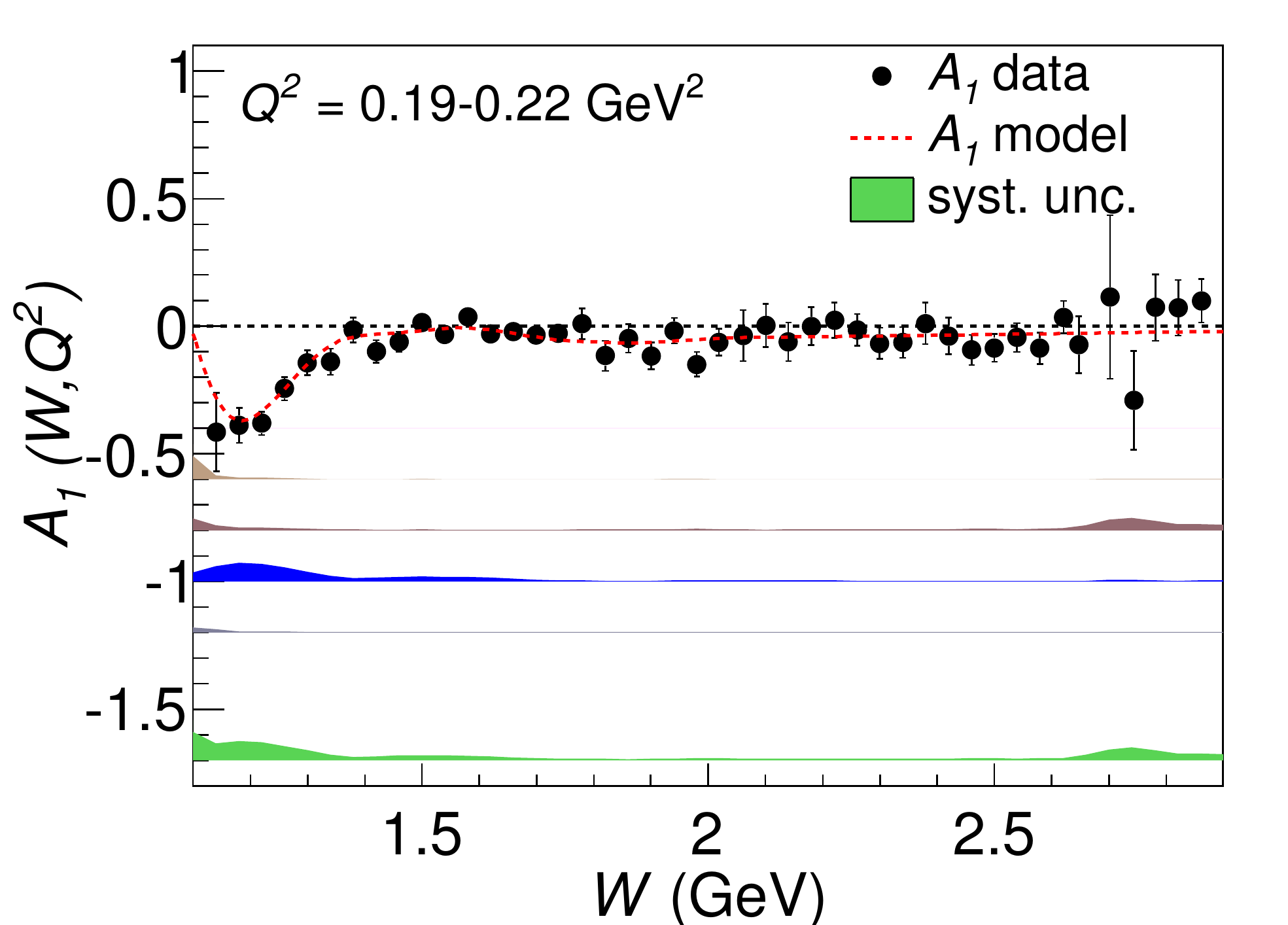} 
  \includegraphics[trim=0cm 2.15cm 0cm 0.4cm,clip=true, width=8cm]{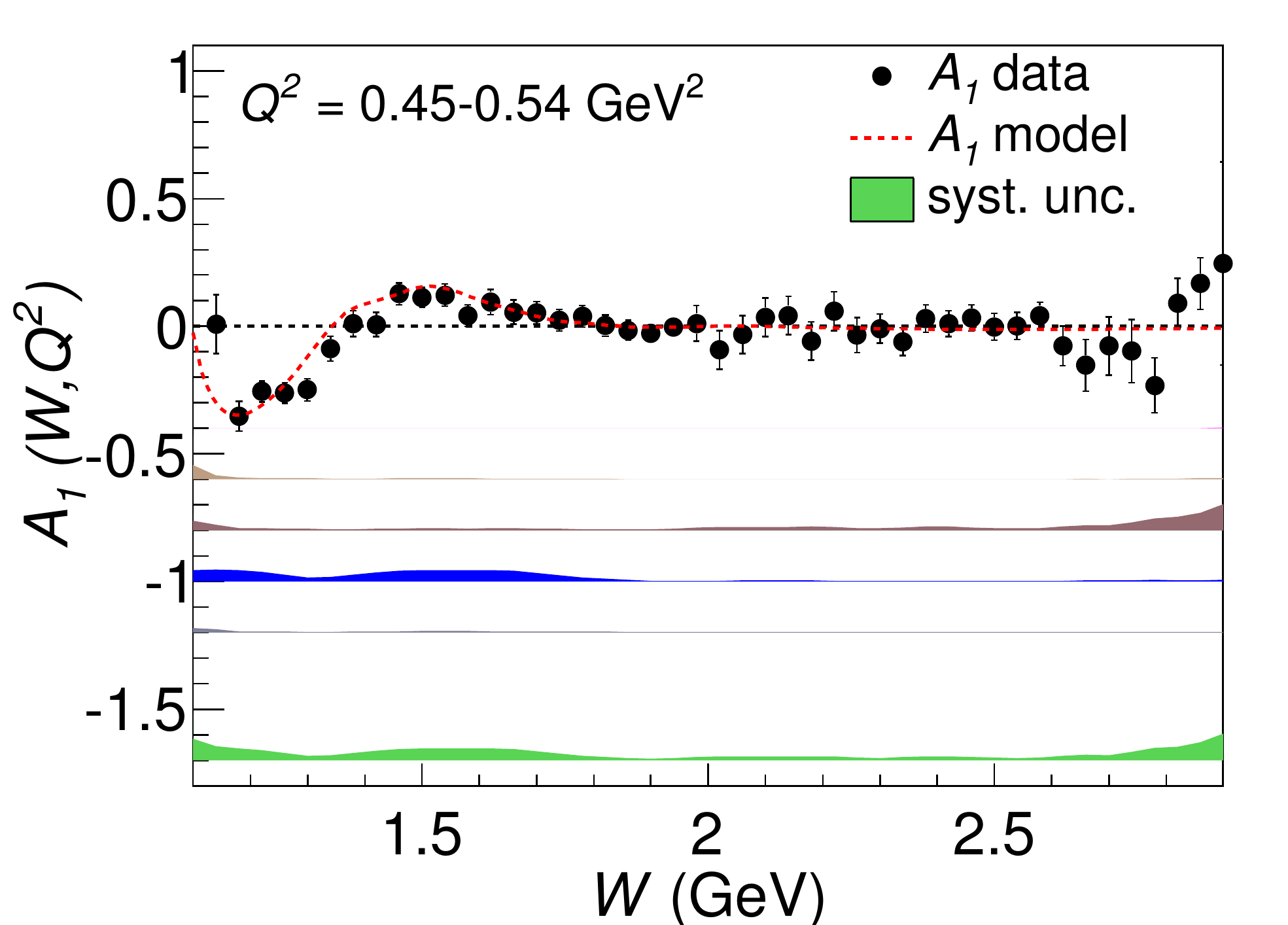} 
  \includegraphics[trim=0cm 0cm   0cm 0.4cm,clip=true, width=8cm] {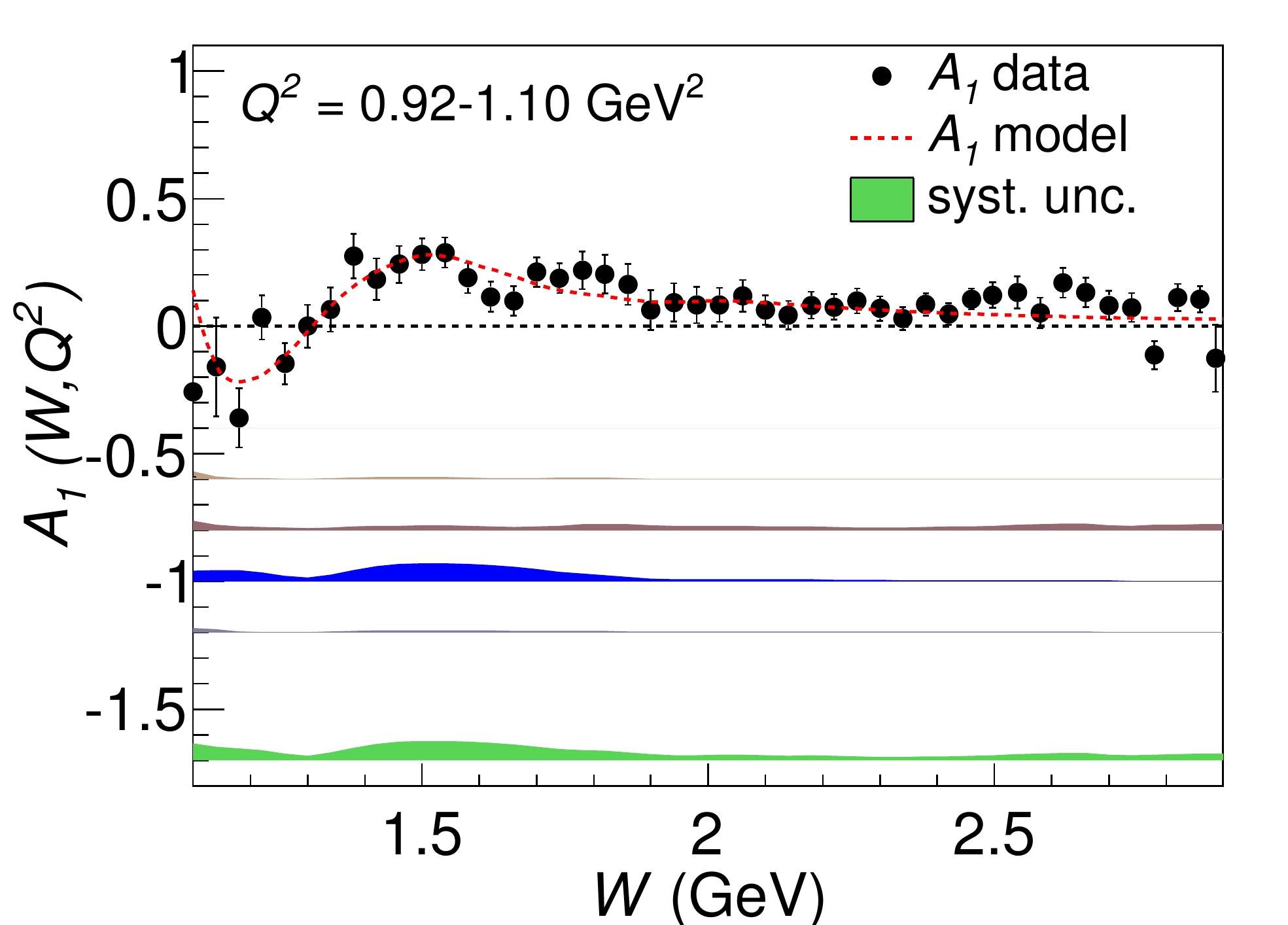} 
\caption[Virtual photon asymmetry $A_1$ for the deuteron versus $W$ for a few $Q^2$ bins.]{
(Color Online) Virtual photon asymmetry $A_1$ for the deuteron versus $W$ for a few $Q^2$ bins:
0.16 GeV$^2 \leq Q^2 \leq 0.19$ GeV$^2$ (top), 0.45 GeV$^2 \leq Q^2 \leq 0.54$ GeV$^2$ (middle) and
1.1 GeV$^2 \leq Q^2 \leq 1.3$ GeV$^2$ (bottom).
The statistical uncertainties are indicated by error bars, while the total systematic uncertainties are indicated by the 
shaded band at the bottom. Again, the individual contributions are shown separately as offset bands:
pion and pair symmetric contamination (-0.4); dilution factor (-0.6); P$_b$P$_t$ (-0.8); models plus radiative corrections (-1.0); and polarized background (-1.2).
}
\label{A1W_ver1:fig1}
\end{figure}

Once $A_1 + \eta A_2$ is calculated, we can extract the virtual photon asymmetry $A_1$, by using a model  for $A_2$  (see Section~\ref{models}). Since $A_1$ depends only on $W$ and $Q^2$, we can combine the results from all beam energies at this stage, again weighted by statistical uncertainties. 
Figure~\ref{A1W_ver1:fig1} shows $A_1(W)$
for three representative $Q^2$ bins
 together with different sources of systematic uncertainties. The 
 uncertainty on $A_2$
(included in the band at -1.0)
 is the dominant contribution to the overall systematic uncertainty (shaded band at the bottom of each panel).

\begin{figure}[htb!]
\centering
\includegraphics[width=9.2cm]{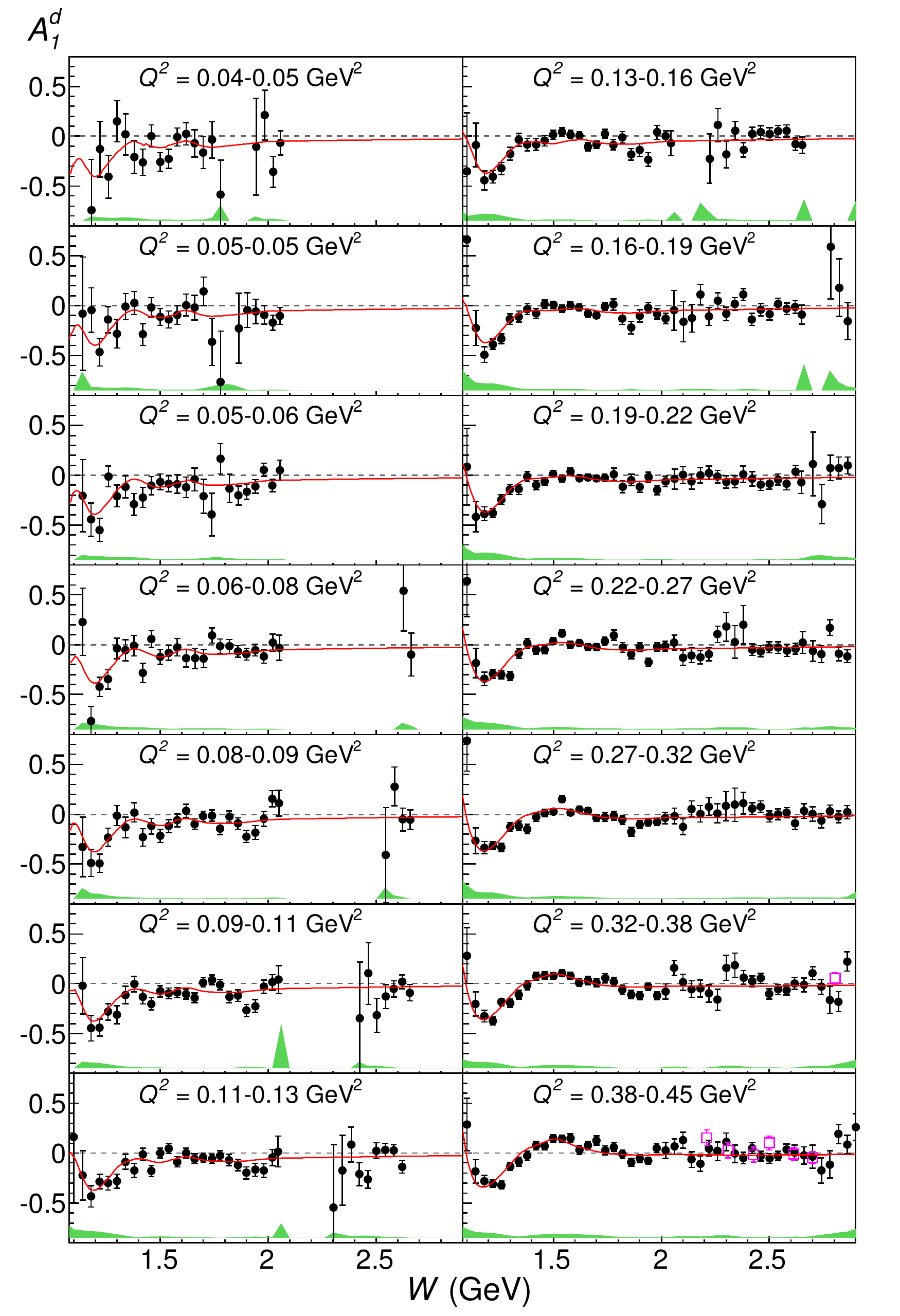}
\caption[$A_{1}$ for the deuteron versus final state invariant mass $W$ for various Q$^{2}$ bins.]
{(Color Online)
$A_{1}$ for the deuteron versus  $W$ for 
our 14 lowest $Q^{2}$ bins. Total systematic uncertainties are shown as shaded area at the bottom of each plot. Our parametrized model is also shown as a red line on each plot. Only the data points with $\sigma_\mathrm{stat} < 0.3$ and 
$\sigma_\mathrm{sys} < 0.2$ are plotted. In addition, we also show data from SLAC 
E143~\cite{Abe:1998wq,Abe:1996ag} (open-magenta squares).
}
\label{A1W_Table1:fig}
\end{figure}

\begin{figure}[htb!]
\centering
\includegraphics[width=9.2cm]{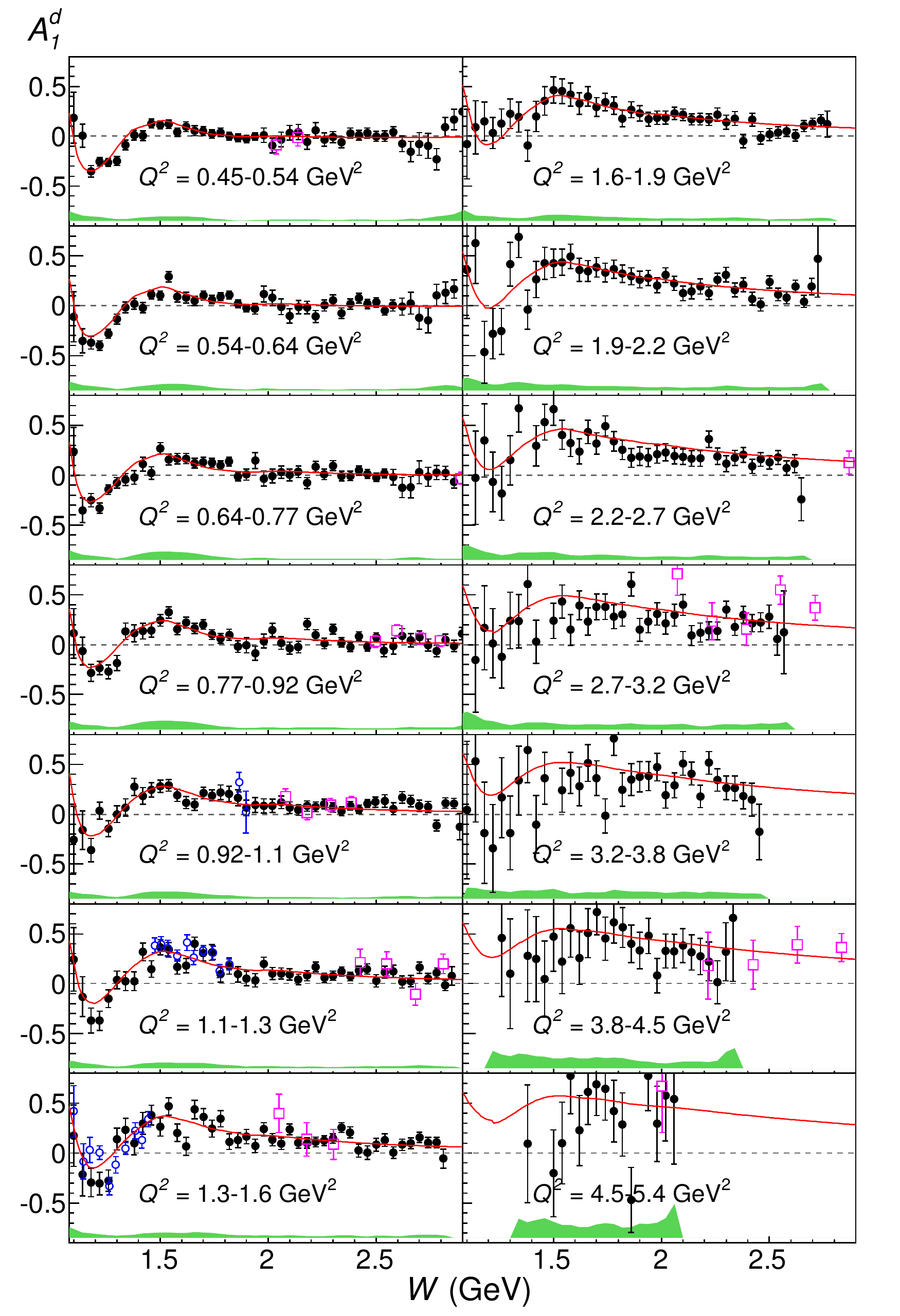}
\caption[$A_{1}$ of the deuteron versus the final state invariant mass $W$ for various Q$^{2}$ bins.]
{(Color Online)
Continuation of Fig. \ref{A1W_Table1:fig} for the remaining $Q^2$ bins. 
In addition to our data and the SLAC data (see above), we also
show the data from the Jefferson Lab RSS 
experiment~\cite{Wesselmann:2006mw,RondonAramayo:2009zz} (blue open circles).}
\label{A1W_Table2:fig}
\end{figure}

Figures~\ref{A1W_Table1:fig} and \ref{A1W_Table2:fig} show $A_1$ versus $W$ for all $Q^2$ bins in our kinematic coverage, as well
as existing data from SLAC 
E143~\cite{Abe:1998wq,Abe:1996ag} and from the Jefferson Lab RSS experiment~\cite{Wesselmann:2006mw,RondonAramayo:2009zz}. Gaps are due to a lack of
 kinematic coverage between the different beam
energies. Data points with very large statistical or systematic uncertainties were omitted from these plots.

At all but the highest $Q^2$, the effect of the $\Delta(1232)3/2^+$ resonance is clearly visible
in the strongly negative values of $A_1$, due to the dominance of the $A_{3/2}$ transition to
this resonance. At our lowest $Q^2$, the asymmetry is in general negative or close to zero, 
which proves that the $A_{3/2}$ transition amplitude is dominant in this region as expected from exclusive pion production. As we go to higher values of $Q^2$ and  $W$, the transition amplitude $A_{1/2}$ leading to resonances such as $N(1520)3/2^-$ and $N(1535)1/2^-$ becomes dominant, as expected from pQCD. At $W > 2$ GeV and larger $Q^2$, the asymmetry continues smoothly from the resonance region into the DIS region where it has been measured by previous experiments to be positive, due to the larger contribution from the proton (with $A_1 > 0$ throughout the measured $x$ range in the DIS region).

\begin{figure}[htb!]
\centering
\subfigure
{
  \includegraphics[width=9.3cm]{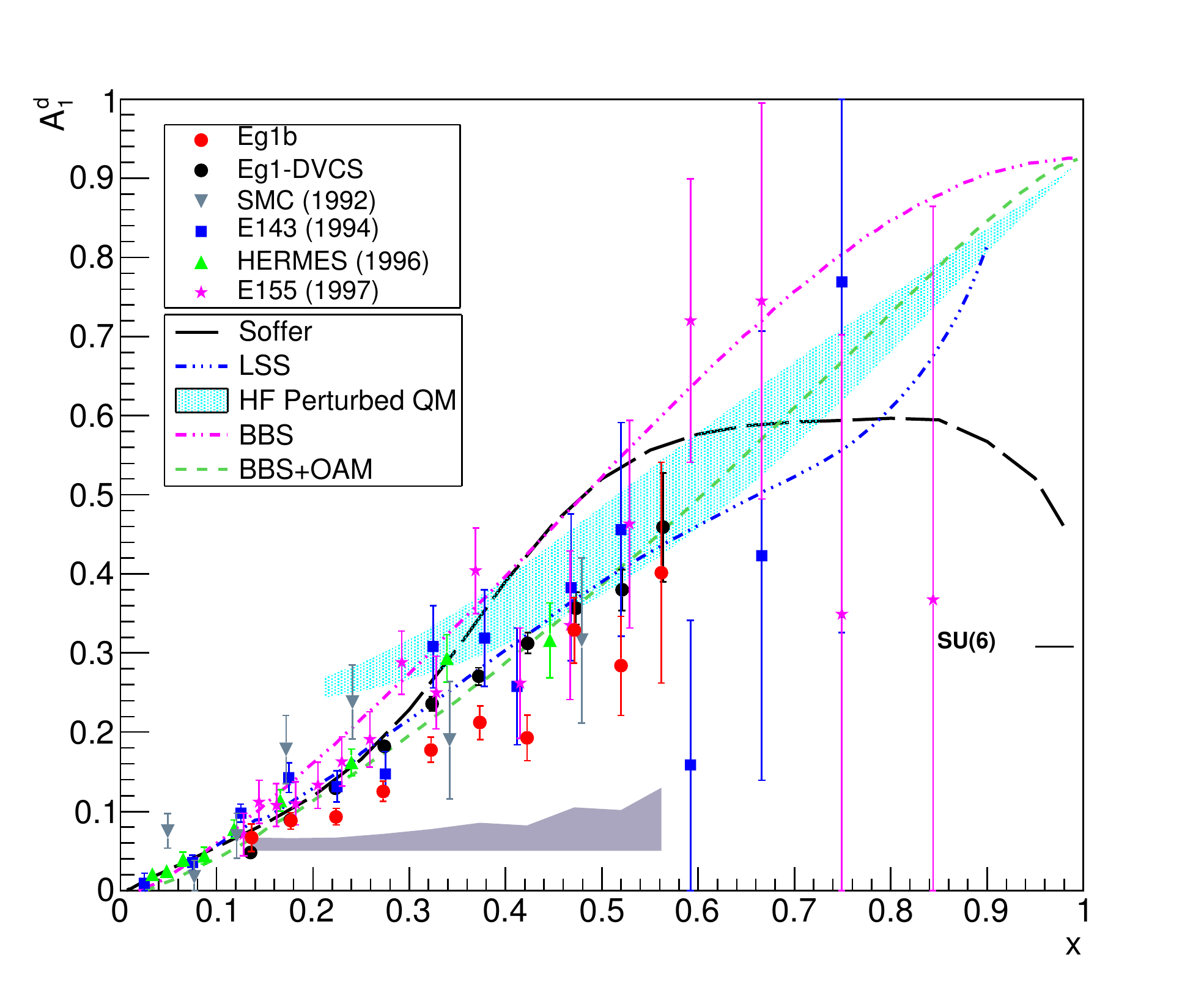} 
}
\caption[$A_1^d$ versus $x_{Bj}$.]{(Color Online)
$A_1^d$ versus $x$  in the DIS region ($Q^2 > 1$ GeV$^2$ and $W > 2$ GeV)
 from EG1b and several other experiments:
 EG1-dvcs at Jefferson Lab~\cite{Prok:2014ltt}, SMC at CERN~\cite{Adeva:1998vv},
 E143 and E155 at SLAC~\cite{Abe:1998wq,Abe:1996ag,Anthony:1999rm} and
 HERMES at DESY~\cite{Airapetian:2006vy}.
 Statistical uncertainties are indicated by error bars, and EG1b systematic uncertainties
by the shaded band at the bottom. Various theoretical predictions and parametrizations are shown as
lines and shaded band, and are discussed in the text.
}
\label{A1intx:fig}
\end{figure}

This trend becomes more apparent if we integrate our data on $A_1$ over the full measured DIS range with
$W > 2$ GeV and $Q^2 > 1$ GeV$^2$ and plot it as a function of the scaling variable $x$.
The behavior of $A_1(x)$ at large $x$ is of high interest to test various models inspired by QCD, as
outlined in Section~\ref{photonabsorption}. Figure~\ref{A1intx:fig} shows this quantity from EG1b together with world data and various models.
We note that our data lie somewhat below most of the world data, which is partially explained by the fact 
that at each point in $x$, they have the lowest average $Q^2$ of all the experiments shown,
implying a more significant impact of scaling violations due to higher twist effects.
In particular,
the new results from EG1-dvcs~\cite{Prok:2014ltt} (also shown in Fig.~\ref{A1intx:fig}) are for 5.9 GeV beam energy and scattering angles above 18$^{\circ}$,
while our data average over 5.7 and 4.2 GeV and scattering angles down to 8$^{\circ}$. 
In addition, systematic differences exist between these two most
precise data sets due to the target polarization, dilution factor, and the different impact from
required model input
for $R$ and $A_2$ at different kinematics.
The corresponding systematic uncertainties are indicated for EG1b by the shaded 
 band at the bottom of the plot.

We also show various predictions based on expectations about the asymptotic value for $A_1^d$ in the limit
$x \rightarrow 1$ 
(see Section~\ref{photonabsorption}). The prediction from a  SU(6)-symmetric quark model is a constant value of 1/3 for 
$A_1^d$ and is indicated by the short horizontal line at the right-hand edge of the plot. A more advanced quark model including hyperfine perturbation through one-gluon exchange~\cite{Isgur:1998yb}
 yields a range of 
possible behaviors at high $x$, as indicated by the shaded (light blue) band. Two different curves (labeled
BBS) are based on pQCD models; one under the assumption of pure quark-hadron helicity 
conservation~\cite{Brodsky:1994kg} and a second one including the effect of a possible non-zero orbital
angular momentum (BBS+OAM~\cite{Avakian:2007xa}). Finally, we show two recent NLO parametrizations
of the world data (by Soffer~\cite{Soffer:2014dba} and by Leader, Stamenov and
 Sidorov -- LSS~\cite{Leader:2014uua}).

We note that, on average, the world data including our own indicate a rise of $A_1^d$ beyond the SU(6)
limit at very large $x$, but much slower than expected from pQCD without the inclusion of orbital angular
momenta. Taking a possible $Q^2$ dependence and systematic uncertainties into account, our data
agree best with the model including orbital angular momenta~\cite{Avakian:2007xa} and are also 
compatible with the lower edge of the range of predictions from the hyperfine-perturbed quark 
model~\cite{Isgur:1998yb}. Overall, no firm conclusion can be drawn yet about the transition of the
down quark polarization from negative values below $x \approx 0.5$ to the limit of +1 expected from
pQCD. A similar conclusion comes from measurements on $^3$He~\cite{Zheng:2003un,Zheng:2004ce}.

\subsection{The spin structure function $g_{1}$}
{\label{sec:g1}}

\begin{figure}[htb]
  \centering
  \includegraphics[width=9cm]{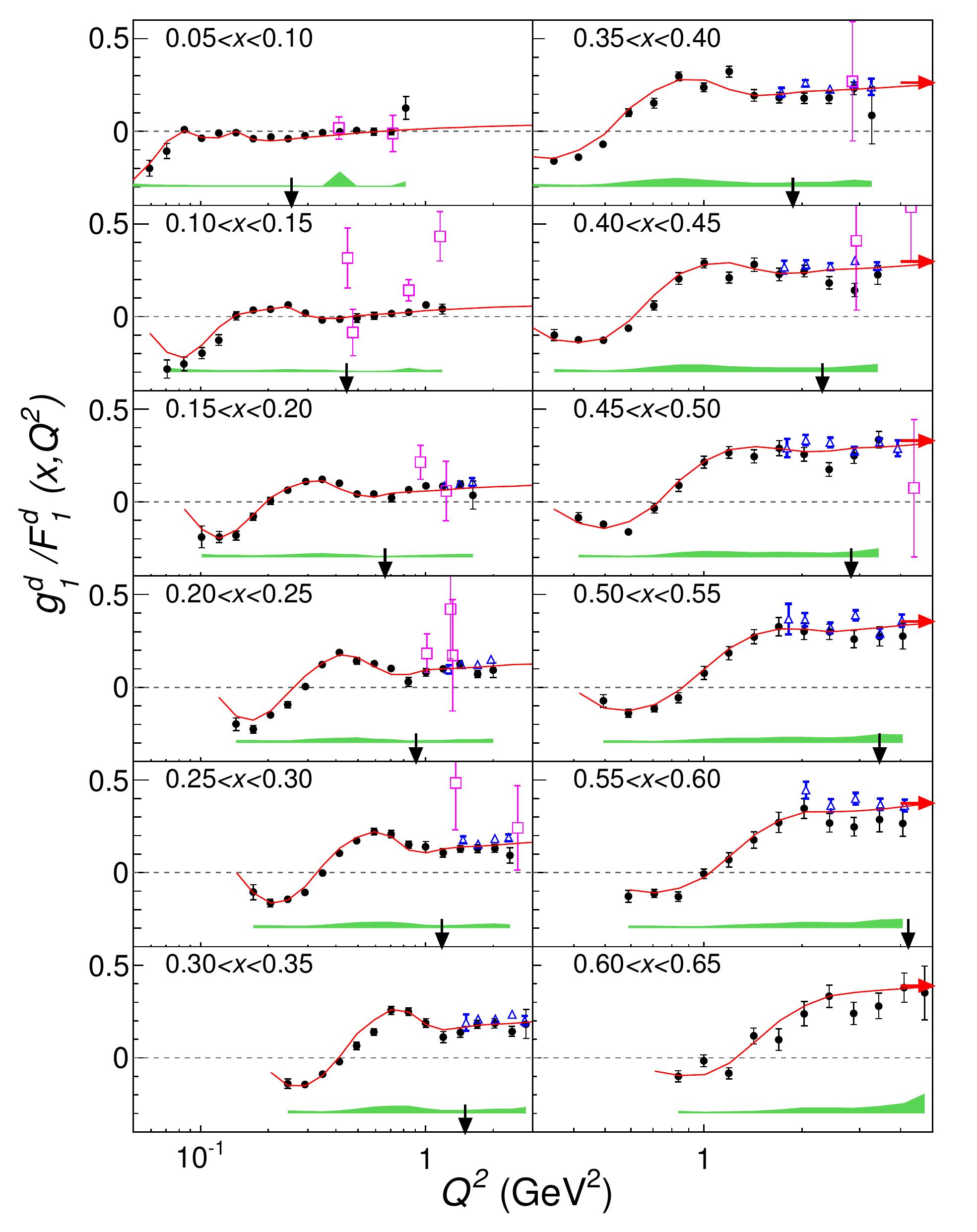}
  \caption[$g_{1}/F_1$ for the deuteron versus Q$^{2}$ and for various  bins in the Bjorken variable $x$ .]
  {(Color Online) The ratio
$g_{1}/F_1$ for the deuteron versus Q$^{2}$ and for various  bins in the Bjorken variable $x$, together with our model shown as the red line on each plot. All data are corrected by our model to center them on the middle
of each $x$ bin.
The shaded area at the bottom of each plot represents the systematic uncertainty. 
Published world data are shown as open-magenta squares (E143~\cite{Abe:1998wq,Abe:1996ag})
and open blue triangles (EG1-dvcs~\cite{Prok:2014ltt}). Arrows
on the x-axis indicate the limit $W=2$ GeV.
The horizontal arrows on the r.h.s of the right panel indicate the results 
for $g_{1}/F_1$  of a recent NLO
analysis of world data~\cite{Leader:2014uua} for our bin centers and $Q^2 = 5$ GeV$^2$. 
  }
  \label{G1F1X_Table1:fig}
\end{figure}

\begin{figure*}[htb]
  \centering
  \includegraphics[width=18cm]{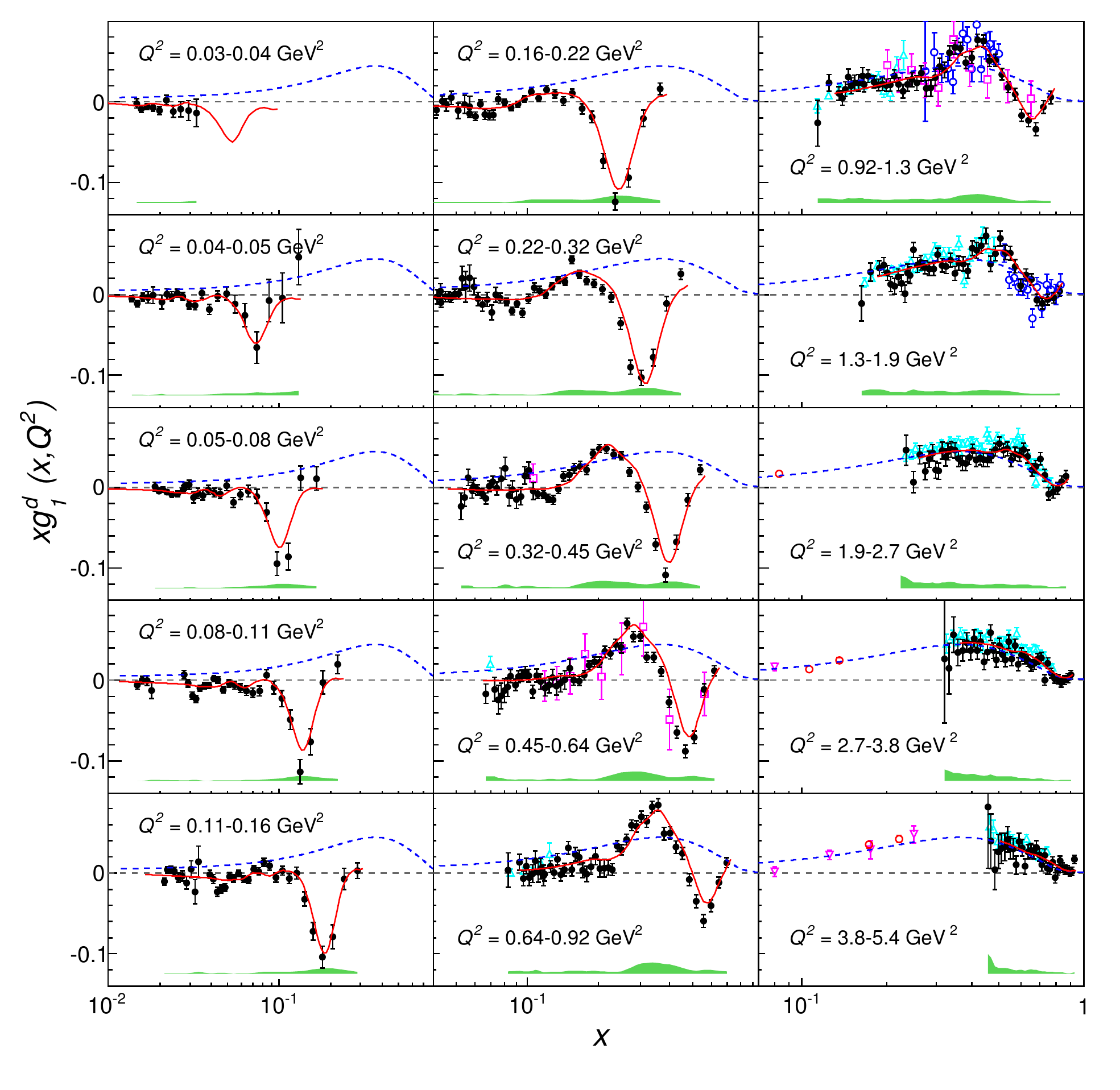}
  \caption[$g_{1}$ for the deuteron versus the Bjorken variable $x$ for various Q$^{2}$ bins.]
  {(Color Online)
The product $x g_{1}$ versus $x$ for all $Q^{2}$ bins, together with our model (red lines). 
The shaded area at the bottom of each plot represents the systematic uncertainty. 
The corresponding DIS parametrization for $Q^2 = 10$ GeV$^2$ is also shown (blue dashed lines). 
 World data are shown for Hermes~\cite{Airapetian:2006vy} (red circles), 
 SLAC E143~\cite{Abe:1998wq,Abe:1996ag} (open-magenta squares), 
 SLAC E155~\cite{Anthony:1999rm} (magenta inverted triangles), 
 RSS~\cite{Wesselmann:2006mw,RondonAramayo:2009zz}  (blue circles), and
 EG1-dvcs~\cite{Prok:2014ltt} (cyan triangles).}
  \label{G1X_Table1:fig}
\end{figure*}


In addition to extracting $A_1$, we can also use the measured asymmetry $A_{||}$ to extract the
spin structure function $g_1^d$ 
according to Eq.~\ref{g1fromApar}. As a first step, we extract the ratio $g_1^d/F_1^d$ which
is less sensitive to various model inputs. Figure~\ref{G1F1X_Table1:fig} shows the resulting
data, plotted for several $x$ bins (all with a bin width of $\Delta x = 0.05$) versus the photon virtuality $Q^2$. Again, we also show world data for the same quantity. Our data agree reasonably
well with those from E143~\cite{Abe:1998wq,Abe:1996ag} within statistical uncertainties,
but are somewhat lower than the very precise data from the recently published 
follow-on experiment EG1-dvcs~\cite{Prok:2014ltt}. However, the difference is consistent
with the (largely uncorrelated) systematic uncertainties of both experiments. 
The $Q^2$ dependence at lower $Q^2$ reflects the effect of nucleon resonances
at $W < 2$ GeV, while beyond this limit (indicated by arrows on the x-axis) this
dependence is mild but still rising, indicating a smooth but not necessarily fast
 transition to the scaling region. We indicate the results for $g_1/F_1$  at $Q^2=5$ GeV$^2$
 from a recent NLO fit of the world data~\cite{Leader:2014uua} for comparison.

We then use models for the unpolarized structure function
$F_1$  (see next section) to convert these ratios to $g_1$.
The results for the product
$x g_1^d$ versus Bjorken $x$ for each of our $Q^2$ bins are presented in Fig. \ref{G1X_Table1:fig}, together with world data. The red curve on each plot comes from  our model. 
At low $Q^2$, $g_1$ is 
strongly affected by  resonance structures, in particular  the $\Delta(1232)$ again being the most prominent one, making $g_1$ negative in this region. When we go to higher $Q^2$, the effect of the resonances diminishes and $g_1$ approaches the smooth DIS curve also shown in 
Fig.~\ref{G1X_Table1:fig} 
as blue dashed line. This can be interpreted as a sign that quark-hadron duality begins to work at these larger
$Q^2 > 1.0$ GeV$^2$. 
However, in the $\Delta(1232)$ region, the data
fall noticeably below the blue line even at $Q^2$ as high as $\approx 1$
GeV$^2$.


In the DIS region ($W > 2$ GeV and $Q^2 > 1$ GeV$^2$), $g_1^d(x)$ can
be used to extract information on the quark helicity contributions to the nucleon spin (see Section~\ref{g1}).
Comparing our data to the higher $Q^2$ data from COMPASS~\cite{Ageev:2007du}
one can extract information on the gluon polarization through DGLAP evolution. 
Including our data for somewhat lower $Q^2$, 
higher twist modifications of the polarized PDFs can be constrained.
Our data are available for such PDF fits, similar to recent fits by the JAM collaboration~\cite{JMO13}
and by Leader et al.~\cite{Leader:2014uua}, 
as well as for future tests of duality.

\subsection{Models}
\label{models}
To extract the physics quantities discussed above  from our data on $A_{||}$, we require models both for 
the unpolarized structure functions $F_1$ and $F_2$ (or, equivalently,
$F_1$ and $R$), as well as for the  asymmetry $A_2$. These models (plus a model for the asymmetry $A_1$)
are also needed to evaluate radiative corrections (Section~\ref{radcorr:sec}) and to extrapolate
our data to small $x$, for the purpose of evaluating moments of $g_1$ (see next section).
For the deuteron case in particular, we need models for both the proton and the neutron, as well
as a prescription for Fermi-smearing. 

We will describe our fit in detail in Ref.~\cite{EG1b_pfin}. Our approach to Fermi-smearing is explained
in Section~\ref{deuteron}. Here, we just summarize our sources of data for the fits to $A_2$ and $A_1$ for the
proton and the neutron.
For the unpolarized structure functions $F_1^{p,n}$ and $R^{p,n}$, we used a recent 
parametrization
of the world data by Bosted and Christy~\cite{Christy:2007ve,Bosted:2007xd}. This 
parametrization fits both DIS and resonance-region 
data with an average precision of 2-5\%, including Jefferson Lab Hall C data on the proton and the deuteron
with very similar kinematics to ours.
Systematic uncertainties due to  these models were calculated by
varying either $F_1$ or $R$ by the average uncertainty of the fit  and recalculating
all quantities of interest.

For the asymmetries in the region $W > 2$ GeV, 
we developed our own phenomenological fit to the world data, including
all DIS results from SLAC, HERA, CERN and  Jefferson Lab
(see Ref.~\cite{Kuhn:2008sy} for a complete list).
In the resonance region, we added data from EG1a~\cite{Fatemi:2003yh,Yun:2002td} in Hall B, 
RSS~\cite{Wesselmann:2006mw} in Hall C and MIT-Bates~\cite{Bates:ref}. 
We also used the data reported here and in~\cite{EG1b_pfin} and iterated the fit after
re-extracting our data using the updated models.
The proton asymmetries
were fit first, followed by a fit to the neutron $A_1$ and $A_2$. For this second part, we
used the rich data set collected on $^3$He at Jefferson Lab (Hall 
A)~\cite{Amarian:2003jy,Kramer:2005qe,Meziani:2004ne,Zheng:2003un,Zheng:2004ce}, 
SLAC~\cite{Anthony:1993uf,Anthony:1996mw,Abe:1997cx,Abe:1997qk},
and HERMES~\cite{Ackerstaff:1997ws,Airapetian:2006vy}, as well as the world
data  on the deuteron, including our own. 
The goodness of the fit ($\chi^2$)
was calculated by comparing the fit functions for neutron asymmetries directly with
neutron results extracted from $^3$He data, as well as comparing the convolution of our
proton and neutron models with corresponding deuteron data.
To anchor our fit of $A_1$ at the photon point, we used data from 
ELSA and MAMI (see, e.g., the summary by Helbing~\cite{Helbing:2006zp}).
As a result, we achieved a consistent fit of proton, deuteron and neutron data over a wide kinematic
range, far exceeding our own kinematic coverage. The overall $\chi^2$ for the fit was
2451 for 3225 degrees of freedom.

Our fit results are shown as curves on most of the plots in this section, and they are generally in very good agreement
with the existing data. 
We developed alternative model fits representing the uncertainty of our fit results in all cases
and estimated the systematic uncertainties on all extracted quantities due to model uncertainties
by replacing the standard fits, one by one, with these alternatives.


\subsection{Moments of $g_{1}$}
{\label{sec:Moments}}

\begin{figure}[htb!]
\centering

\subfigure
{
  \includegraphics[trim=0cm 0cm 0cm 0cm, clip=true, width=8cm]{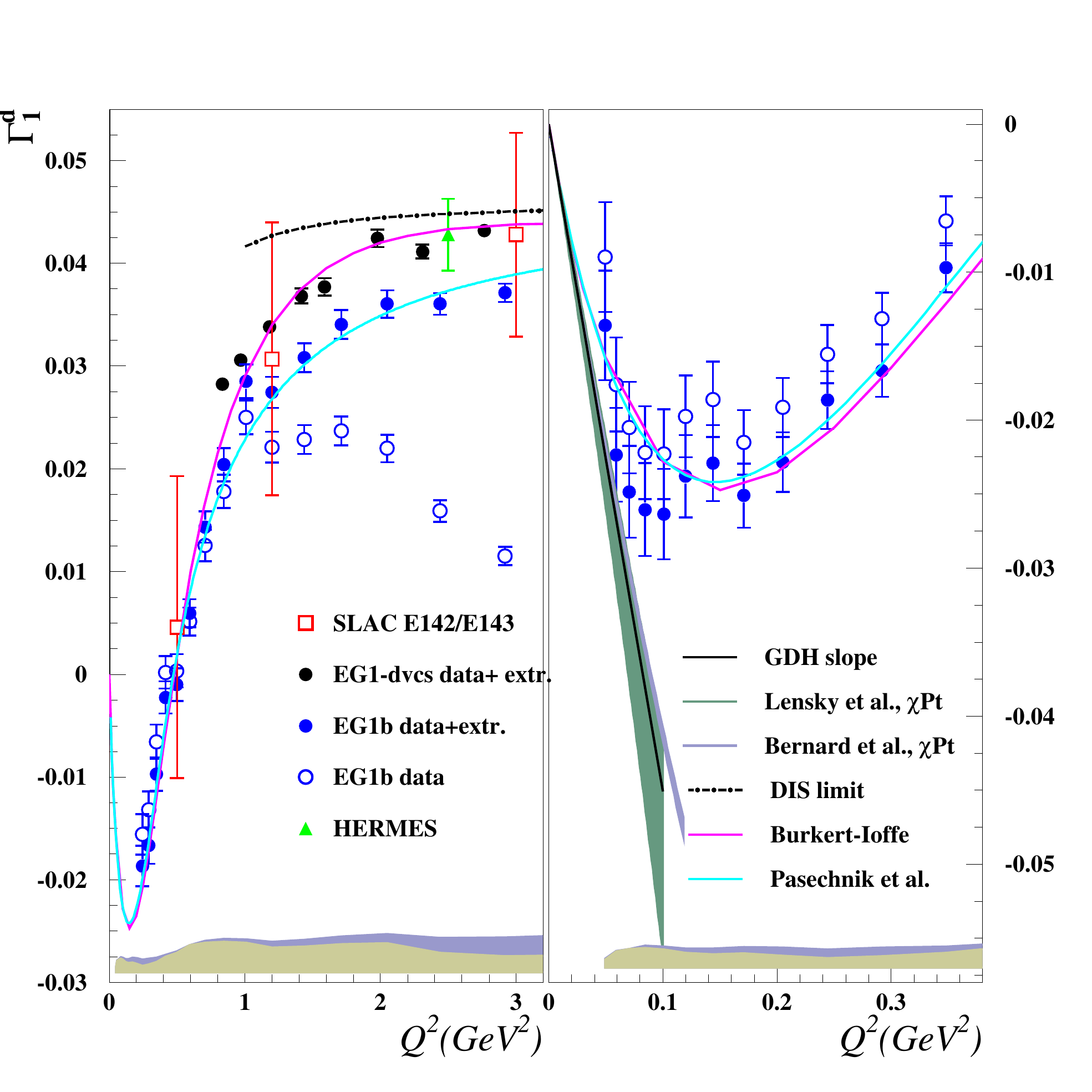} 
}
\caption[$\Gamma_1$ for the deuteron versus $Q^2$ from data and data+model.]{(Color Online)
$\Gamma_1$ for the deuteron versus $Q^2$ from our data only (hollow blue circles) and 
from data plus model (full blue circles), including the extrapolation to the unmeasured kinematics. The 
left-hand side shows  the full $Q^2$ range
(leaving out our data for $Q^2 < 0.3$ GeV$^2$, to avoid clutter)
 and the right-hand side focuses on the 
small-$Q^2$ region.
The systematic uncertainty is shown at the bottom of the plot, for data only (light beige shaded area in 
the foreground) and for combined data and model (blue shade in the background).  
Corresponding results from 
SLAC E143~\cite{Abe:1998wq,Abe:1996ag}, HERMES~\cite{Airapetian:2006vy}
 and  EG1-dvcs~\cite{Prok:2014ltt} are
shown, as well as several predictions (explained in the text).
}
\label{Gamma1_data_full:fig}
\end{figure}

From our data, we determined several moments of spin structure functions.
We evaluated those moments 
for each of our standard $Q^2$ bins in two parts. For $W$ regions where we have good data
(with reasonably small statistical uncertainties), we summed directly over these data (binned
in 10 MeV bins in $W$), multiplied by the corresponding bin width in $x$ and the required power of $x$. 
We avoided the region below
$W = 1.15$ GeV, where radiative effects and the quasi-elastic contribution overwhelm the data.  
The upper end of the integration range can go up to $W = 3$ GeV, depending
on the $Q^2$ bin. 
 The resulting values of the integral over the kinematic region covered by our data are shown as the 
open (blue) circles in Fig.~\ref{Gamma1_data_full:fig}, and the properly propagated
 systematic uncertainty in the measured region
 is shown as the light beige band. Note that all moments are calculated
per nucleon (i.e., divided by 2 for the two nucleons in deuterium), following common practice. However, we
do {\em not} correct for the deuteron D-state or any other nuclear effects.

We integrate our model for $g_1^d$ (without any quasi-elastic contributions)
over the region $1.08$ GeV $\le W \le 1.15$ GeV in order
to estimate this small part of the full moment~\footnote{We exclude the (quasi-)elastic region $W < 1.08$ GeV, following common convention, since the quasi-elastic peak would overwhelm the integrals at small $Q^2$. }.
Occasionally, there are gaps in our $W$ coverage from different beam
energies, especially at low $Q^2$ (see, e.g., Fig.~\ref{A1W_Table1:fig}). These gaps are also filled by integrating
the model instead. Finally, we integrate the model from  the lower $x$ limit of our highest $W$ bin (for each $Q^2$)
down to $x = 0.001$. This contribution becomes most important at high $Q^2$ and for the lowest (first) moment.
We limit ourselves to this minimum $x$ value because there are no reliable data at lower $x$, and 
our model becomes unconstrained and rather uncertain
below $x = 0.001$. While it is likely that there is no significant contribution
below this limit~\footnote{The contribution from $x < 0.001$ is  most certainly negligible for the higher moments.}, 
we prefer to quote our results as moments from $x=0.001$ to $x_{max}$, where
\begin{equation}
x_{max} = \frac{Q^2}{W_{min}^2 - M^2+Q^2}
\end{equation}
and $W_{min} = 1.08$ GeV.
The values of the full integral for the first moment are shown in Fig.~\ref{Gamma1_data_full:fig} as the filled
(blue) data points and the full systematic uncertainty due to the additional model uncertainty in the unmeasured region is indicated by the  wider blue band behind the beige one.
We also show published world data on the first moment in the same $Q^2$ range.
Our data are again in reasonable agreement with the world data (within statistical
uncertainties) except for being slightly below the data from EG1-dvcs~\cite{Prok:2014ltt}
as mentioned before; again, the difference is consistent with the systematic uncertainty
on both experiments. At $Q^2 < 0.8$ GeV$^2$, ours are the only high-precision data
available so far, extending down to $Q^2= 0.05$ GeV$^2$, where they can be used
to test effective theories like Chiral Perturbation Theory ($\chi$PT).

 We compare our results with several theoretical predictions and parametrizations
 in Fig.~\ref{Gamma1_data_full:fig}.
The black dashed-dotted curve indicates
the extrapolation from the DIS limit using pQCD corrections up 
to third order in $\alpha_s$, assuming the asymptotic value for the moment from recent
publications by COMPASS~\cite{Alexakhin:2006vx} and HERMES~\cite{Airapetian:2006vy}.
We also show two parametrizations that connect the DIS limit with the real photon point.
One parametrization, by Burkert  {\it et al.}~\cite{BurkertIoffe2}
(upper magenta curve), combines an estimate
of the integral in the resonance region with a smooth function connecting the photon
point, constrained by the Gerasimov-Drell-Hearn (GDH) sum 
rule~\cite{Gerasimov:1965et,Drell:1966jv}, with the asymptotic limit.
The second, by Pasechnik {\it et al.}~\cite{Pasechnik:2010fg} (light blue line), 
includes both higher-twist terms at large $Q^2$ and a chiral-like expansion at the
photon point within the framework of an analytic perturbation theory (APT) which
 has been fit to available data, including previous (partial) results from
EG1b~\cite{Prok:2008ev}. Both parametrizations do a remarkably good job
describing the world data on the first moment over the full range of $Q^2$.

\begin{figure}[htb]
\centering
 \includegraphics[width=8.5cm]{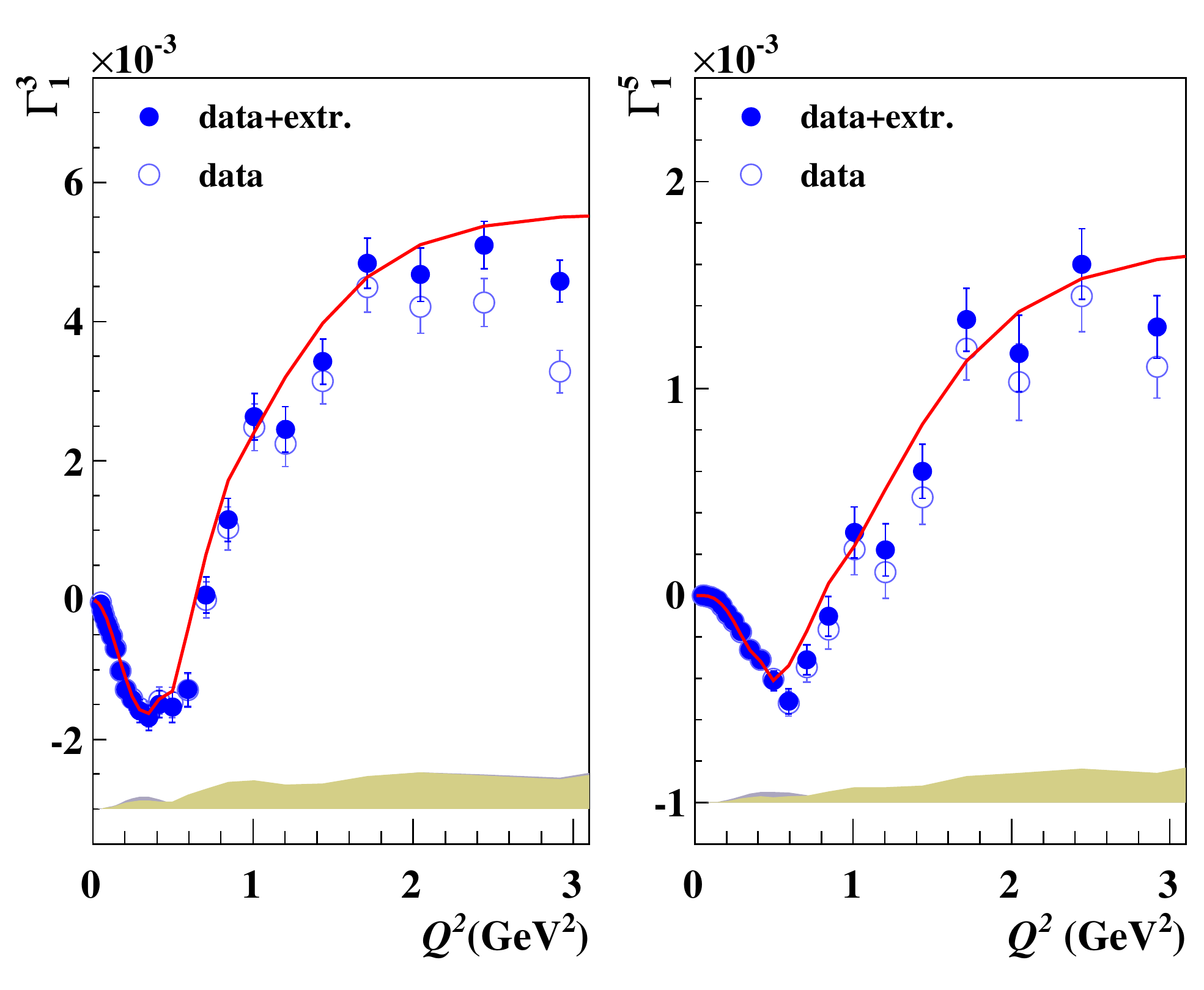}
\caption[$\Gamma_1^3$ and $\Gamma_1^5$ versus $Q^2$.]{
(Color Online) Higher moments of $g_1^d$ extracted from the EG1b data versus $Q^2$.
The third moment for the deuteron, $\Gamma_1^3$, is shown on the left, 
and the fifth moment, $\Gamma_1^5$, on the right. The open squares were calculated with no model contribution while the filled squares include model input for the kinematic regions with no available data. 
The total systematic uncertainty is shown by the blue (experimental only) and
black (experimental plus extrapolation) shaded areas.
}
\label{GammaHigher_data:fig}
\end{figure}

We also show several predictions for the low--$Q^2$ behavior of $\Gamma_1$ 
on the right-hand side of
Fig.~\ref{Gamma1_data_full:fig}, including the slope at $Q^2 = 0$ from the 
GDH sum rule~\cite{Gerasimov:1965et,Drell:1966jv} (solid black line)
and its extensions from two recent chiral perturbation theory calculations.
The first one, by Bernard {\it et al.}~\cite{Bernard2013aa} (narrow dark grey band
on r.h.s.).
is an expansion up to third order
with explicit inclusion of $\Delta(1232)3/2^+$ isobar degrees of freedom. The second, by
Lensky {\it et al.}~\cite{Lensky:2014dda,Lensky:2014fza} (wider dark green band),
uses Baryon $\chi$PT
including pion, nucleon and $\Delta(1232)$ degrees of freedom
to calculate all moments
in next-to-leading order (NLO).
Both predictions are close to the GDH limit and show little sign of the 
observed deviation of the
data towards less negative values as $Q^2$ increases; however, they agree with
our lowest three points  $Q^2 < 0.08$ GeV$^2$
within their statistical and systematic uncertainties.

 The higher moments $\Gamma_1^3$ and $\Gamma_1^5$ are also calculated in the same way with appropriate powers $n=3,5$ (see Section~\ref{formalism}). Fig. \ref{GammaHigher_data:fig} shows the results for the third moment $\Gamma_1^3$ and the fifth moment $\Gamma_1^5$ of $g_1$ 
 from the EG1b data. These moments
 are useful for the extraction of higher twist matrix elements, e.g., the third moment is 
 directly related to the matrix element $a_2$ within the Operator Product Expansion.

\begin{figure}[htb]
\centering
  \includegraphics[width=7cm]{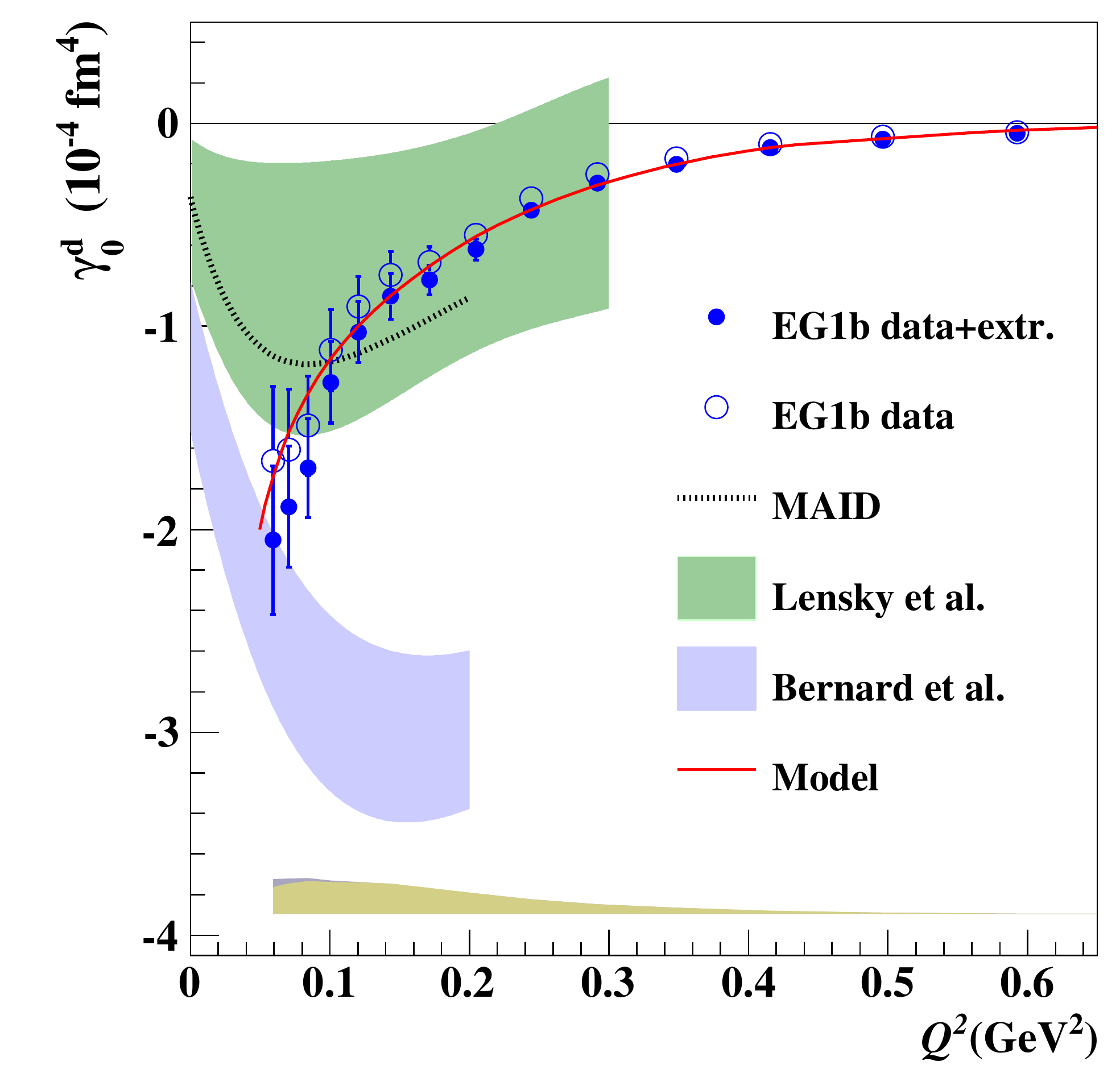} 
\caption[Forward Spin Polarizability ($\gamma_0$) versus $Q^2$.]{
(Color Online) Forward spin polarizability ($\gamma_0$) for the deuteron  versus $Q^2$. The open squares represent the result using only data and the solid black circles are data plus model results. The shaded area close to the $x$-axis is the total systematic uncertainty (blue for experimental only and black with extrapolation
uncertainty included). 
Our model  is   shown as a red solid line. Our results are compared to three
 $\chi$PT calculations (see text) and the MAID parametrization~\cite{Kamalov:2001yi}
  for single pion production.
}
\label{gamma0:fig}
\end{figure}

To calculate the extended spin polarizability $\gamma_0$, we integrate the product of $A_1 F_1$ 
instead of $g_1$, weighted with $x^2$. The result is multiplied by $16 M^2 (\hbar c)^4 \alpha/Q^6$
to convert to [$10^{-4}$ fm$^{4}$], in agreement with the definition for real photons.
Fig. \ref{gamma0:fig} shows our result for the forward spin polarizability $\gamma_0$ for the deuteron.
We compare them again to the $\chi$PT calculations by 
Lensky {\it et al.}~\cite{Lensky:2014dda,Lensky:2014fza} (upper yellow band)
and by Bernard {\it et al.}~\cite{Bernard2013aa} (lower light blue band) as well as
an evaluation of single pion production data by the MAID collaboration~\cite{Kamalov:2001yi}.
The $\chi$PT calculations do not quite reproduce the trend of the data at low $Q^2$.

\subsection{Neutron spin structure functions}
{\label{sec:Neutron}}
\begin{figure}[htb]
  \centering
  \includegraphics[width=8.5cm]{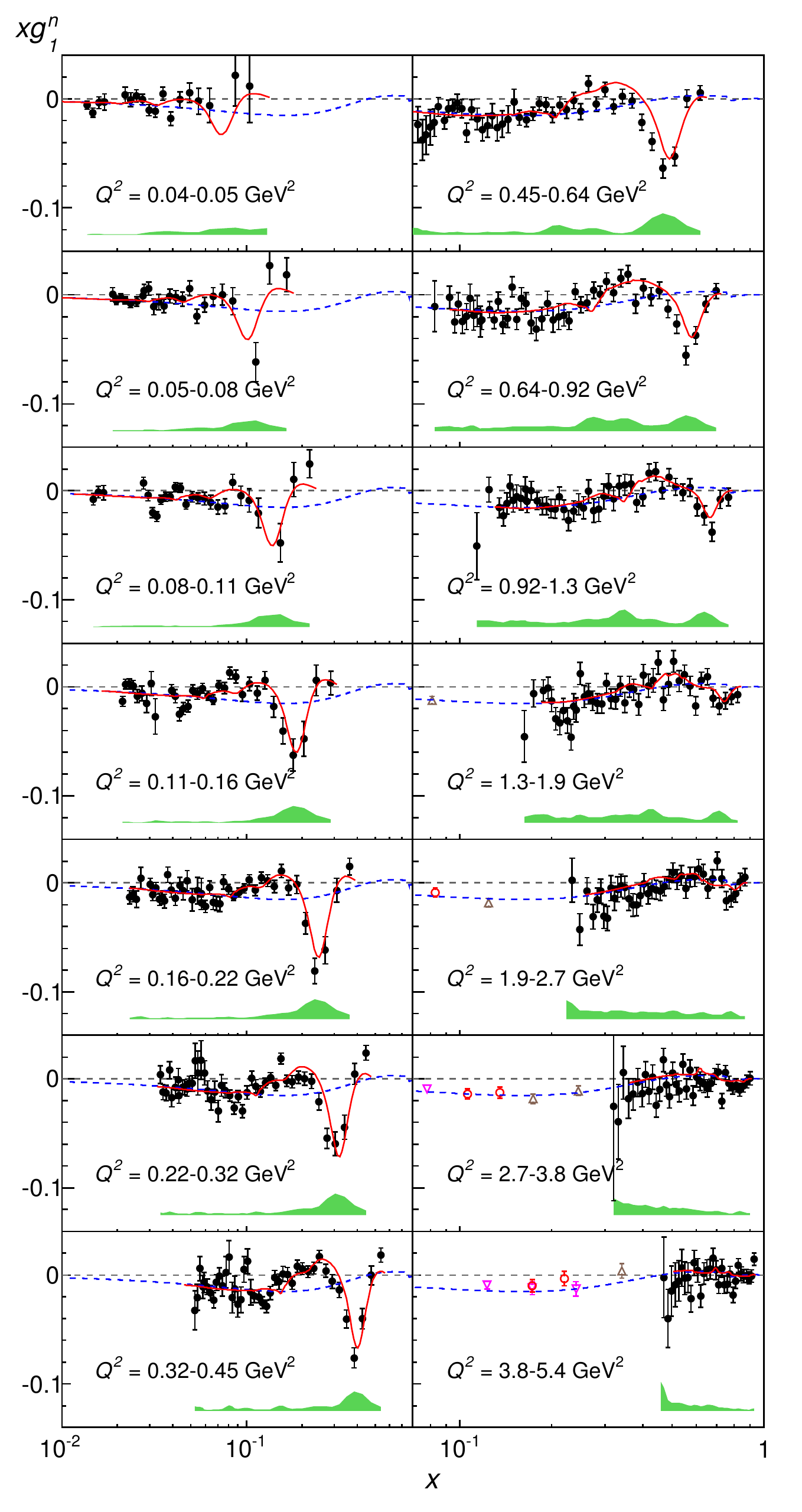}
  \caption[$g_{1}$ for the bound neutron versus the Bjorken variable $x$ for various Q$^{2}$ bins.]
  {(Color online) Our results for the spin structure function
$x g_{1}$ of the bound neutron, extracted in the impulse approximation framework of Ref.~\cite{Kahn:2008nq} versus the Bjorken variable $x$ for all (combined) Q$^{2}$ bins
(filled circles). Our
model is shown as red lines on each plot, and the asymptotic form of $g_1(x)$ in the DIS
region is shown as dashed blue lines. The shaded area at the bottom of each plot represents the systematic uncertainty. Additional data from other experiments are shown as well: E154~\cite{Abe:1997cx} (magenta inverted triangles), HERMES~\cite{Ackerstaff:1997ws,Airapetian:2006vy} (red circles) and 
E142~\cite{Anthony:1996mw} (brown triangles). 
  }
  \label{G1X_n_Table1:fig}
\end{figure}

\begin{figure}[htb!]
  \centering
  \includegraphics[width=9cm]{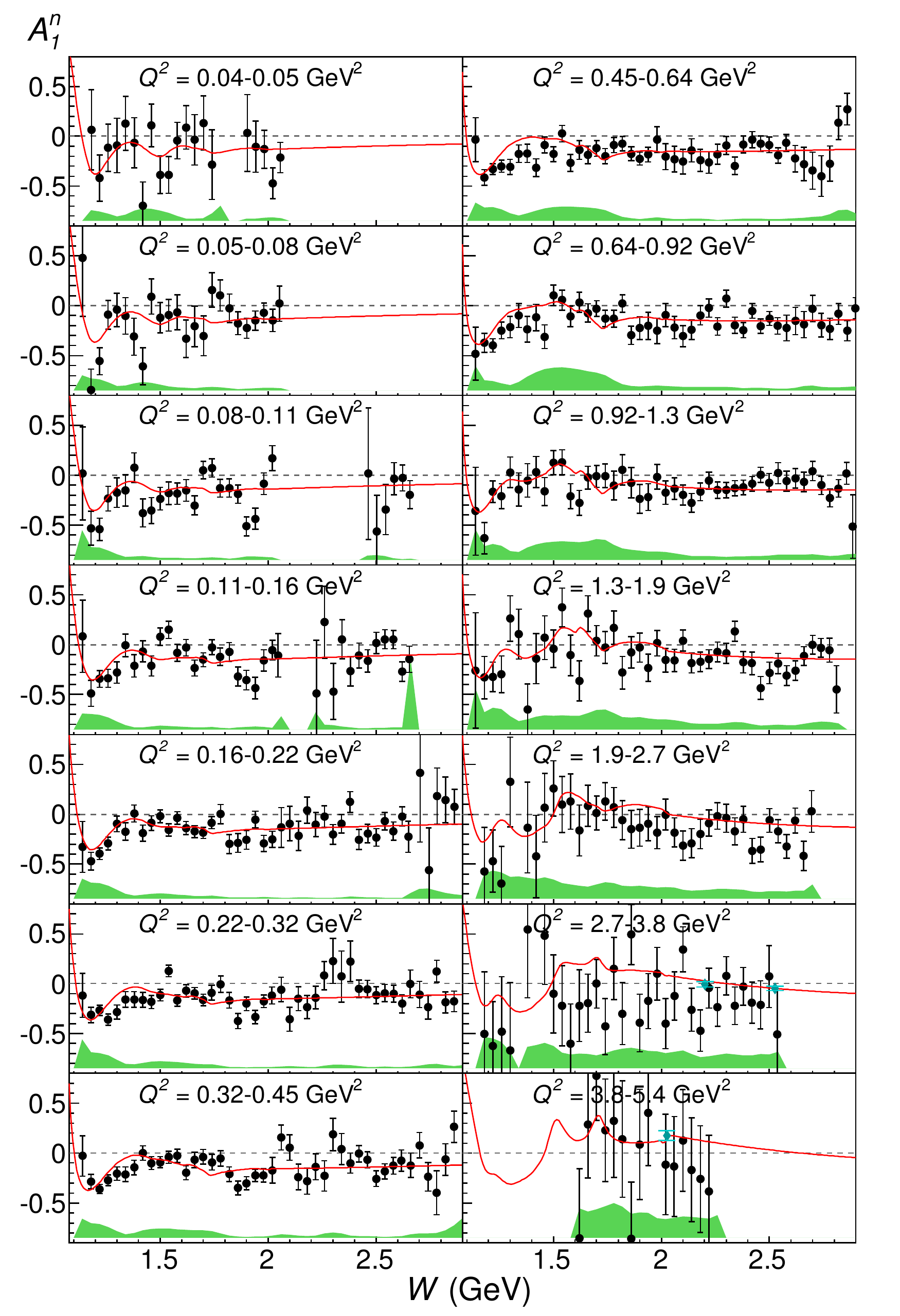}
  \caption[$A_{1}$ for the neutron versus the final state invariant mass $W$ for various Q$^{2}$ bins.]
  {(Color Online)
$A_{1}$ for the bound neutron, extracted from our results for $g_{1}^n$
  (see Fig.~\ref{G1X_n_Table1:fig}), versus   $W$ for our combined Q$^{2}$ bins. Systematic uncertainties are shown as shaded area at the bottom of each plot. Our parametrized model is also shown as a red line on each plot. Only the data points with $\sigma_\mathrm{stat} < 0.6$ and $\sigma_\mathrm{sys} < 0.2$ are plotted. The cyan diamonds indicate data from measurements on $^3$He~\cite{Zheng:2003un,Zheng:2004ce}. 
  }
  \label{A1W_n_Table1:fig}
\end{figure}

Although many data sets exist for spin structure functions of the (bound) neutron in the
deep inelastic (DIS) region, no un-integrated results have been published in the region $W < 2$ GeV
of the nucleon resonances. This is due to the difficulty of reliably extracting neutron information
from measurements that  have to use nuclear targets, as explained in
Section~\ref{deuteron}. As discussed in that section, we have attempted, for the first
time, to combine our deuteron data with our proton fit (Section~\ref{models}) and an impulse
approximation folding prescription
to  access information on the neutron in a model-dependent way, see Figs.~\ref{G1X_n_Table1:fig} and
\ref{A1W_n_Table1:fig}.

Our method relies on the folding prescription by Kahn {\it et al.}~\cite{Kahn:2008nq}
which describes deuteron structure functions in terms of those of the proton and the neutron. 
We used this prescription in our fit for the asymmetries $A_1^n$ and $A_2^n$ for the
neutron as described in Section~\ref{models}. In particular, for any set of fit parameters,
we calculate both $g_1^n, g_2^n$ and $g_1^p, g_2^p$, combine them 
(following Ref.~\cite{Kahn:2008nq}) and compare directly
to the measured $g_1^d$. The parameters are optimized until the best possible
agreement (smallest $\chi^2$) is achieved. 

Our extraction of  the $g_1^n$ data points shown in Fig.~\ref{G1X_n_Table1:fig}
follows a slightly different procedure than that described in Ref.~\cite{Kahn:2008nq}, but is similar
to their ``additive'' method: We assume that any difference between the measured
and the calculated
$g_1^d$  is solely due to a corresponding discrepancy in $g_1^n$ at that
specific kinematic point. Given that to first approximation 
\begin{equation}
g_1^d \approx (1- 1.5 P_D)(g_1^n + g_1^p) 
\end{equation}
we then calculate 
\begin{equation}
g_1^n (\mathrm{meas}) = g_1^n (\mathrm{model}) +
 \frac{g_1^d(\mathrm{meas}) - g_1^d(\mathrm{model})}{1- 1.5 P_D} ,
\end{equation}
with $P_D \approx 0.05$.
This method has the advantage that it is stable (as opposed to trying to invert the folding)
and that it leads to a straightforward propagation of statistical uncertainties: 
\begin{equation}
\sigma(g_1^n) (\mathrm{meas}) = 
 \frac{\sigma(g_1^d)(\mathrm{meas}) }{1- 1.5 P_D} .
\end{equation}
The systematic uncertainties are evaluated in the same fashion as in all previous cases
(see Section~\ref{syserr:sec}), by
varying one model input or experimental parameter in sequence and propagating the
variation to the final result for $g_1^n (\mathrm{meas})$, then adding all of these
variations in quadrature.
The final results are shown in Fig.~\ref{G1X_n_Table1:fig}, together with world data
at higher $W$ and both our model parametrization (red line) and the DIS limit
at $Q^2 = 10$ GeV$^2$ (blue dashed line). We  combined our
standard $Q^2$ bins pairwise for clarity of presentation.

\begin{figure}[htb]
  \centering
  \includegraphics[width=9cm]{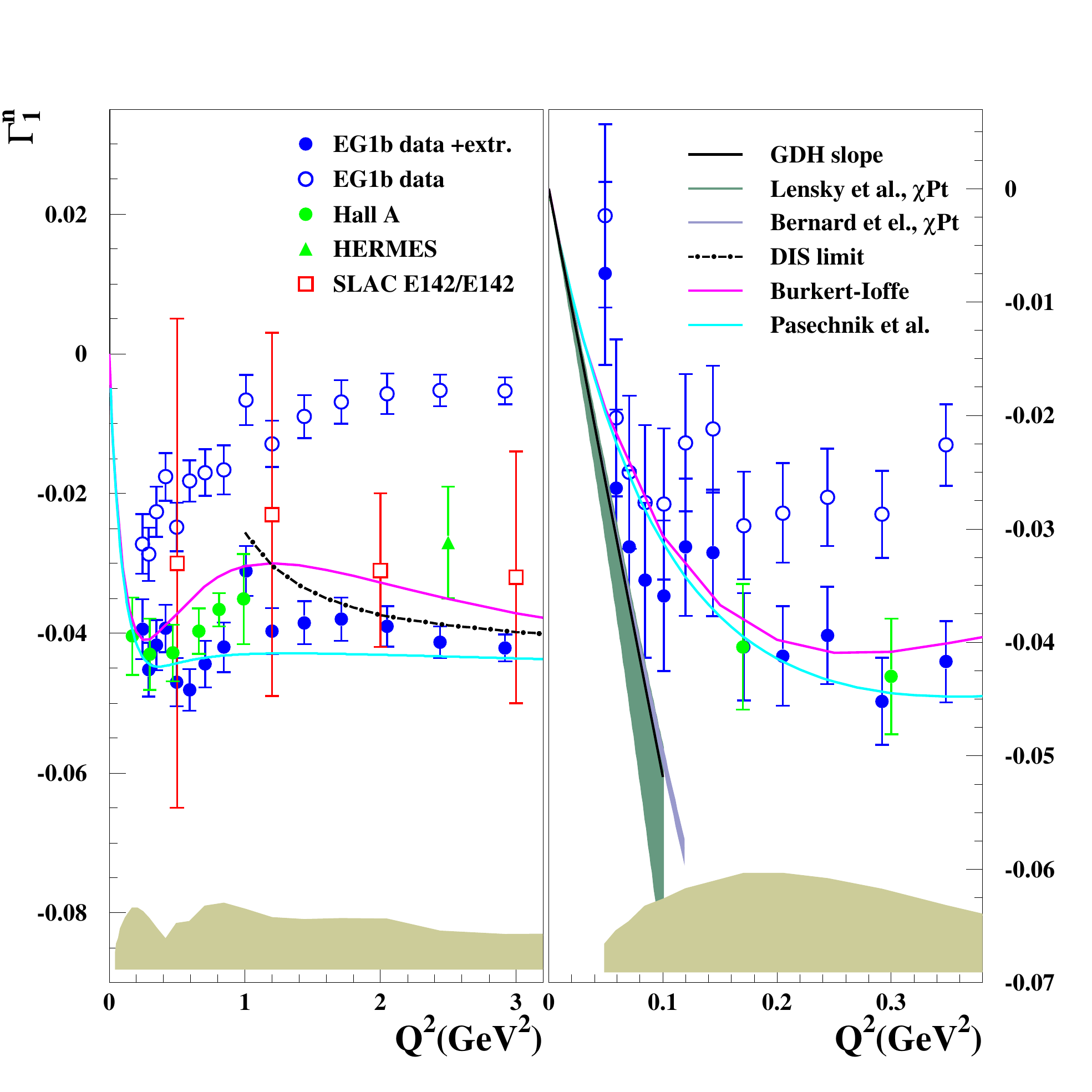}
  \caption
  {(Color online) $\Gamma_1$ for the neutron versus $Q^2$ from data only (open blue circles) and data plus model (full blue circles), including the extrapolation to the unmeasured kinematics. 
  Also shown are phenomenological calculations from Pasechnik {\it et al.}~\cite{Pasechnik:2010fg}
  (lower light blue line)
   and Burkert  {\it et al.}~\cite{BurkertIoffe2} (upper magenta line), together with the $\chi$PT results from 
   Lensky {\it et al.}~\cite{Lensky:2014dda,Lensky:2014fza} (wider dark green band) and 
   Bernard {\it et al.}~\cite{Bernard2013aa} (thin grey band). The GDH slope (black solid line) and pQCD prediction (black dotted line) are also shown. The right-hand side plot is a magnification of
    the low $Q^2$ region
   (which is omitted from the l.h.s.). 
   Systematic uncertainties of our data are shown as shaded areas at the bottom of the plot. 
   Results from other experiments are also shown, with statistical and systematic uncertainties (added
   in quadrature) reflected in their total error bars.  }
  \label{Gam1_n_Table1:fig}
\end{figure}


As a next step, we can then convert the results for $g_1^n$ into the virtual photon
asymmetry $A_1^n$, by using our models for $F_1^n$ and $A_2^n$. The results are shown in
Fig.~\ref{A1W_n_Table1:fig}. Overall, the agreement of the extracted  results with our
 model is quite good, except at the highest $Q^2$ where our data seem to lie systematically
lower (a trend that can already be observed in the corresponding deuteron data, see
Fig.~\ref{A1W_Table2:fig}). We direct the attention of the reader to the additional data
points plotted in the last two $Q^2$ bins (cyan diamonds); these are the results from
the Hall-A experiment on $^3$He~\cite{Zheng:2003un,Zheng:2004ce}
 at the highest attainable $x$ in the DIS region. These data are consistent with our own, but
 with significantly smaller statistical uncertainties. However, no such data have been published
for any of the lower
 $Q^2$ bins.

\begin{figure}[htb]
\centering
  \includegraphics[width=7cm]{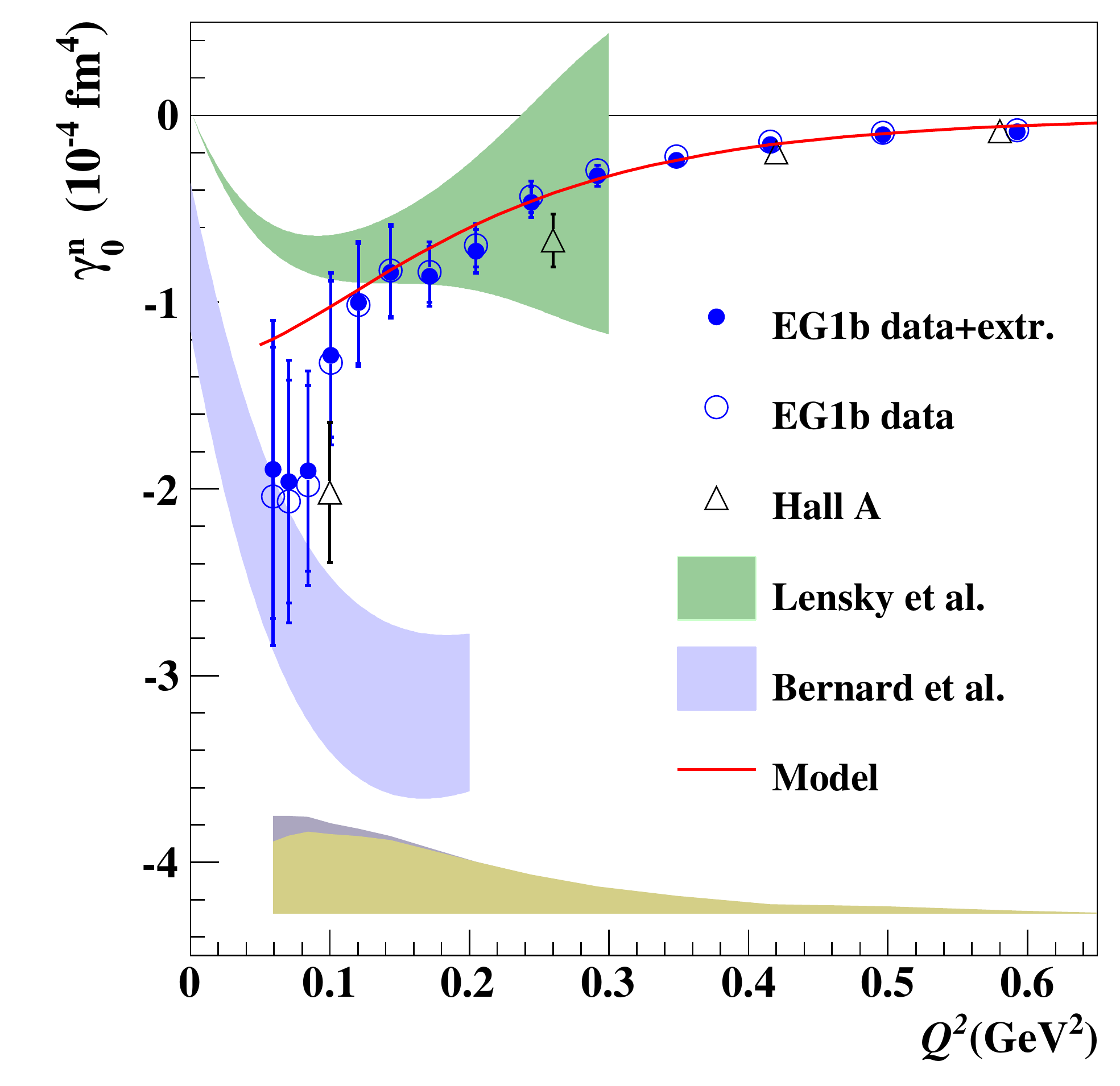} 
\caption[Forward Spin Polarizability ($\gamma_0$) versus $Q^2$.]{
(Color Online) Forward spin polarizability $\gamma_0$ for the neutron versus $Q^2$. 
The open squares represent the result using only data and the solid black circles are data plus model results. The shaded area close to the $x$-axis is the total systematic uncertainty (blue for
experimental data only and black including the extrapolation).
Our model  is also shown as a red solid line. Our results are compared to three
 $\chi$PT calculations (see text) and to the $^3$He data from Hall A~\cite{Amarian:2003jy}.
}
\label{gamma0_n:fig}
\end{figure}

As a final step, we once again form various moments of the neutron spin
structure functions (see Figs.~\ref{Gam1_n_Table1:fig} and \ref{gamma0_n:fig}).
While the advantage of using deuterium as a proxy for the neutron (namely, its
much smaller average nucleon momenta and therefore less severe kinematic
smearing) is less clear in this case (since the moments integrate over all kinematics
anyway), it is still instructive to compare our results to those using a $^3$He target
as a source of polarized neutrons~\cite{Amarian:2003jy}. Again, we find good agreement 
between these two experiments using different effective neutron targets and
with very different systematic uncertainties. We note that the neutron data are also
well described by the two parametrizations~\cite{Pasechnik:2010fg,BurkertIoffe2},
while they approach the GDH limit above (but marginally compatible with) the Chiral
Perturbation calculations~\cite{Lensky:2014dda,Bernard2013aa}.

Figure~\ref{gamma0_n:fig} shows the forward spin polarizability for the bound neutron from our
data, again compared to data from the $^3$He experiment in Hall A~\cite{Amarian:2003jy}. 
The agreement at the lowest $Q^2$ is excellent, and our data extend to slightly lower $Q^2$.
Once again, they show a general agreement with the order of magnitude predicted by $\chi$PT
while exhibiting a distinctly different shape with $Q^2$. 

\section{CONCLUSION}\label{s6}
In summary, we present the final analysis of the most extensive data set on 
the spin structure functions $A_1$ and $g_1$ of the deuteron in the valence and resonance region. 
The data cover two orders of magnitude
in squared momentum transfer, $0.05 \leq Q^2 \leq 5$ GeV$^2$, connecting the region
of hadronic degrees of freedom and effective theories like $\chi PT$ near the photon point
with the regime where pQCD is applicable. Our data give more detailed insight in the
inclusive response of the deuteron in the resonance region and how it connects
with the DIS limit. They can constrain NLO fits (including higher twist corrections) of 
spin structure functions extracting polarized PDFs, and they shed new light on the 
 valence quark structure of the nucleon at large $x$.They can be used to study
  quark-hadron duality and to extract matrix elements in the framework of the Operator Product Expansion.
  To facilitate such analyses, we are providing the raw data (with minimum
  theoretical bias) through the CLAS experimental database~\cite{DataBase} as well as 
  Supplemental Material for this paper~\cite{SupMat}.
 
 We use our data on the deuteron, together with a detailed fit of the corresponding  proton data,
 to extract bound neutron spin structure functions, using a convolution model and ignoring
 FSI and other binding effects. These results 
 give information, for the first time, on inclusive neutron spin structure in the resonance region
 $W < 2$ GeV. They can also be used
 to cross check the results from $^3$He targets at high $x$. We find general agreement between
 the data from these rather different approaches, within the relatively larger statistical 
 uncertainties of our data set. On the other hand, our data cover a larger range in $Q^2$ and $W$.
 
 Our data allow precise determinations of moments of $g_{1}^d$ (and $g_{1}^n$) as a function
 of $Q^2$, which can
 be used to test the approach to the GDH sum rule limit, $\chi PT$ and phenomenological
 models, and to extract matrix elements in the framework of the  Operator Product Expansion. We find
 that $\chi PT$ describes our results for $\Gamma_1$ only up to very moderate $Q^2 \approx 0.08$
 GeV$^2$ (within our statistical and systematic uncertainties), while there is only
 rough agreement in magnitude between  $\chi PT$ and our data 
 for the forward spin polarizability $\gamma_0$.
Finally, we would like to refer the reader to a recent analysis of the world data~\cite{Deur:2014aa}, 
including the data
presented here, to study the first moment of the difference $g_1^p - g_1^n$ and its $Q^2$--dependence
to extract Operator Product Expansion matrix elements.
 
 Further data will come
from the analysis of the EG4 experiment with CLAS, which will extend the kinematic 
coverage of the present data set to even lower $Q^2$ for a more rigorous test of $\chi PT$.
Additional information on
the structure functions $g_2$ and $A_2$ is forthcoming once experiment ``SANE'' in Hall C
and experiment  ``g2p'' in Hall A have been analyzed. 
Finally, a complete mapping of spin structure functions in the valence quark region, out to the
highest possible $x$, is one of the cornerstones of the program with the energy-upgraded
12 GeV accelerator at Jefferson Lab.

\section*{Acknowledgments}
This material is based upon work supported by the U.S. Department of Energy, Office of Science, Office of Nuclear Physics under contracts DE-AC05-06OR23177, DE-FG02-96ER40960 and other contracts.
Jefferson Science Associates (JSA) operates the Thomas Jefferson National Accelerator Facility for the United States Department of Energy.
We would like to acknowledge the outstanding efforts of the staff
of the Accelerator and the Physics Divisions at Jefferson Lab that made
this experiment possible.  This work was supported in part by
the U.S.  National
Science Foundation,
the Italian Instituto Nazionale di Fisica Nucleare, 
the French Centre
National de la Recherche Scientifique, the French Commissariat \`{a}
l'Energie Atomique,  the Emmy Noether grant from the Deutsche Forschungs
Gemeinschaft,
the Scottish Universities Physics Alliance (SUPA),
 the United Kingdom's Science and Technology Facilities Council,
the Chilean Comisi\'on Nacional de Investigaci\'on Cient\'ifica y Tecnol\'ogica (CONICYT),
 and the National Research Foundation of Korea.


\end{document}